\documentclass[11pt]{article}

\usepackage{graphicx}
\usepackage{amssymb,amsmath}
\usepackage{multirow,cite}

\numberwithin{equation}{section}

\newcommand{\II}{\relax{\rm I\kern-.18em I}}

\newcommand{\rra}[1]{ \setbox1=\hbox{\kern10pt${#1}$\kern10pt}
  \,\vbox{\offinterlineskip\hbox to\wd1{\hfil\copy1\hfil}
    \kern 3pt\hbox to\wd1{\rightarrowfill}}\,}
\newcommand{\Bket}[1]{|\kern-.16em |#1\rangle\kern-.24em \rangle}
\newcommand{\Bovlap}[2]{\langle\kern-.24em \langle#1|\kern-.16em |#2\rangle\kern-.24em \rangle}

\title{\Large{\bf{D-branes in Toroidal Orbifolds and Mirror Symmetry}}}
\author{Eleonora Dell'Aquila}

\date{}
\begin{document}

\begin{flushright}
NSF-KITP-05-110
\end{flushright}
\vspace{0.5cm}
\begin{center}
{\Large\bf{D-branes in Toroidal Orbifolds and Mirror Symmetry}}\vspace{0.7cm}\\
{\large Eleonora Dell'Aquila}
\end{center}
\vspace{0.1cm}
\begin{center}
Kavli Institute for Theoretical Physics\\
University of California, Santa Barbara, CA 93106-4030 \\
\vspace{0.2cm}
NHETC, Department of Physics, Rutgers University \\
136, Frelinghuysen Road, Piscataway NJ 08854 \\
\end{center}
\vspace{0.1cm}
\begin{abstract}
We study D-branes extended in $T^2/\mathbb{Z}_4$ using the mirror description as a tensor product of minimal models. 
We describe branes in the mirror both as boundary states in minimal models and as matrix factorizations in the corresponding Landau-Ginzburg model. 
We isolate a minimal set of branes and  give a geometric interpretation of these as D1-branes constrained to the orbifold fixed points. This picture is supported both by spacetime arguments and by the explicit construction of the boundary states, adapting the known results for rational boundary states in the minimal models.  
Similar techniques apply to a larger class of toroidal orbifolds.
\end{abstract}

\section{Introduction and outline}

There is a finite number of distinct two-dimensional toroidal orbifolds, classified by the crystallographic space groups of the plane \cite{Dijkgraaf, Wendland}. An orbifold by a discrete group of order greater than two can only be realized if the torus is defined by either a square or a hexagonal lattice. The basic examples are $T^2/\mathbb{Z}_4$ (square lattice), $T^2/\mathbb{Z}_3$ and $T^2/\mathbb{Z}_6$ (hexagonal lattice). 
As string backgrounds, these  are rigid: since it is not possible to deform the shape of the torus (complex structure) without breaking the discrete symmetry that is gauged, such deformations are not part of the moduli space. As a consequence of this fact, mirror symmetry relates these toroidal orbifolds  to nongeometric backgrounds, that can be described as asymmetric orbifolds - just applying T-duality - or more interestingly as Landau-Ginzburg models (see for example \cite{Chun, Martinec}).

In this paper we will specialize to the $T^2/\mathbb{Z}_4$ orbifold, but the general discussion applies to the other cases as well. It was shown in \cite{Chun} that the $T^2/\mathbb{Z}_4$ orbifold at the self-dual radius is equivalent, at the level of the worldsheet CFT's, to the tensor product of minimal models $A_2\otimes A_2\,$. This CFT has also a description as a Landau-Ginzburg model with superpotential $W_{LG}=Y_1^4+Y_2^4\,$ \cite{Vafa,Martinec}. Using the Landau-Ginzburg description one can extend the equivalence to the full moduli space of the two models, identifying polynomial deformations of the LG superpotential with the K\"ahler deformations of the orbifold. The aim of this paper is to extend this correspondence to the open string sector and in particular to use the mirror description to study  D-branes extended in the $T^2/\mathbb{Z}_4$ orbifold. 

In the past two years there has  been much progress in working with branes in Landau-Ginzburg models \cite{K-LiAG, Brunner, K-Li, K-LiMM, Lazaroiu, Ashok, Hori, HLL, AshokII, HerbstII, HoriII, HerbstIII, BrunnerII, Walcher, Ezhuthachan, BGpermutation, permutationII, BrunnerIII,Caviezel}, in the framework of matrix factorizations. This approach was first proposed by M. Kontsevich and  was later introduced in the physics literature in \cite{K-Li,K-LiAG,Brunner}. For the LG models that describe (orbifolds of) tensor products of $\mathcal{N}=2$ minimal models \cite{Zamolodchikov,Vafa,Witten,Martinec}, a useful correspondence has been established between LG matrix factorizations and boundary states \cite{Brunner, Ashok, K-LiMM,BGpermutation, permutationII, BrunnerIII}. It has also been possible in certain cases to give a geometric interpretation to the LG branes, either by moving to a different point in moduli space \cite{Ashok} or via mirror symmetry \cite{BrunnerII}. 

In the Landau-Ginzburg setup it is natural to look for a set of fundamental branes, such that all other branes in the model (more precisely we will be considering branes that preserve B-type worldsheet supersymmetry) can be constructed as bound state of these.  Combining the information obtained from the LG analysis with the known results for boundary states in minimal models (especially \cite{RS,Recknagel}) - and the relation between the two - we are able to give a geometric interpretation to this set of branes as D1-branes constrained to pass through any two of the orbifold fixed points, much like fractional branes. 
This provides a very explicit example of mirror symmetry, in addition to allowing us to learn more about the properties of branes that are interesting in their own right.

As mentioned above, a similar analysis could be repeated for other models. In addition to the main examples of $\mathbb{Z}_{3,4,6}$ orbifolds, for which the equivalence to tensor products of minimal models is described very explicitly in \cite{Chun},  one can consider modding out these by various (quantum) discrete symmetries, completing the list of  $T^2$ orbifolds. These models will have a description as Landau-Ginzburg orbifolds. Higher dimensional toroidal orbifolds can also be constructed from these basic building blocks. For example, it is very easy to generalize the computations of this paper to the spacetime supersymmetric $T^4/\mathbb{Z}_4$ orbifold limit of $K3\,$; it would be interesting to study the open string theory on the branes extended in the orbifold directions for this case \cite{us}. Also, for non-supersymmetric backgrounds (such as the one considered in this paper), studying the dynamics of branes extended in the orbifold might shed some light on the evolution of the compact spacetime under closed string tachyon condensation, following the line of reasoning of \cite{APS}.  
We expect that the interplay between Landau-Ginzburg and CFT techniques could prove useful in analysing all these diverse models. 

\subsection*{Organization}

The plan of the paper is the following. Section \ref{CFT} is a review of the equivalence between the $T^2/\mathbb{Z}_4$ orbifold and the tensor product of minimal models $A_2\otimes A_2\,$, rewritten from the point of view of mirror symmetry. The remaining sections contain an analysis of the branes in these models. In section \ref{LGbranes} the formalism of matrix factorizations is used to give a partial classification of B-branes in the Landau-Ginzburg model that describes the $A_2\otimes A_2$ CFT. From this analysis we derive some properties of  the branes that are used, in section \ref{Abranes}, to give a geometric interpretation of the LG branes as D1-branes on $T^2/\mathbb{Z}_4\,$. The arguments  given in this section are based on spacetime intuition and are not sufficient to give a full description of the branes as boundary states. In order to do that, in section \ref{MMBS} we make contact with the description of the same branes as minimal model boundary states. This analysis is more rigorous and allows us to obtain the full boundary state description of the A-branes on $T^2/\mathbb{Z}_4\,$, complementing the more heuristic (but more transparent) discussion of section \ref{Abranes}.
The details of the map between the $T^2/\mathbb{Z}_4$ and $A_2\otimes A_2\,$ CFT's, which are used throughout the paper and especially in section \ref{MMBS}, are collected in an appendix. 

\subsection*{Acknowledgments}

I would like to thank my advisor, Emanuel Diaconescu, for suggesting the problem and for his help and suggestions.  I am very thankful to Sujay Ashok for his advice and encouragement and for his comments at all stages of this work. 
 I am also grateful to Ilka Brunner, Benjamin Doyon, Emiliano Imeroni, Gregory Moore, Christian R\"omelsberger and Jan Troost  for helpful discussions and correspondence. I would like to thank KITP for its hospitality in the past four months.
This research was supported in part by the National Science Foundation under Grant No. PHY99-07949.

\section{Equivalence of $T^2/\mathbb{Z}_4\,$ and $A_2 \otimes A_2 $}\label{CFT}

The model we consider is a $T^2/\mathbb{Z}_4\,$ orbifold with $\mathcal{N}=2$ worldsheet supersymmetry. The coordinates on the torus are $(x^1, x^2)\sim (x^1+2\pi R, x^2 +2\pi R)\,$, with $R=\sqrt{2}$ in units with $\alpha'=2\,$. It will also be convenient to introduce complex coordinates $x^\pm=(x^1\pm i x^2)/\sqrt{2}\,$. In terms of these the $\mathbb{Z}_4$ action is
$x^\pm \to \omega^{\pm 1} x^\pm\,$,
with $\omega=e^{2\pi i/4}$, and similarly on the fermions. It was shown explicitly in \cite{Chun} that this model is equivalent to the tensor product of two $\mathcal{N}=2$ minimal models of the $A_k$-series, with $k=2\,$. In the conventions that we are following, the minimal model $A_{k=2}$ has central charge $c=\frac{3k}{k+2}=\frac{3}{2}$ and has a description as a Landau-Ginzburg model with superpotential $W_{LG}=Y^{k+2}=Y^4\,$. Some facts will be summarized below and further details can be found in the appendix. This equivalence is in fact mirror symmetry and the goal of this section will be to describe the correspondence between the $(a,c)$ chiral ring of the orbifold and the $(c,c)$ chiral ring of the minimal model, in the LG description. This will be useful later to establish a correspondence between the branes in the two models by comparing their couplings to the bulk fields.

The identification between the $T^2/\mathbb{Z}_4$ orbifold and the $A_2\otimes A_2$ conformal field theory is established comparing the superconformal algebra, the spectrum and the OPE's  of the two theories \cite{Chun}.
We adopt the notation
\begin{align}\label{Tsca}
&T= -\frac{1}{2}(\partial x^i \partial x^i + \psi^i\partial \psi^i)\qquad  &&\bar{T}= -\frac{1}{2}(\bar{\partial} x^i \bar{\partial} x^i +\bar{\psi}^i\bar{\partial} \bar{\psi}^i)\nonumber\\
&G^{\pm}= \psi^{\pm} \partial x^{\mp}\equiv e^{\pm iB_L}\partial x^{\mp}\qquad &&\bar{G}^{\pm}= \bar{\psi}^{\pm} \bar{\partial} x^{\mp}\equiv e^{\pm iB_R}\bar{\partial} x^{\mp}\\
&Q=i\partial B\qquad  &&\bar{Q}=-i\bar{\partial} B\nonumber
\end{align}
for the superconformal algebra of the theory with target space $T^2\,$. At the selfdual radius the torus has an enhanced $(SU(2)\times SU(2))^2$ symmetry (left- and right-moving), but the $\mathbb{Z}_4$ projection reduces the enhanced symmetry to $(U(1))^2 \,$. Hence the chiral algebra of the $T^2/\mathbb{Z}_4$ orbifold contains, in addition to the currents \eqref{Tsca}, the holomorphic $U(1)$ current
\begin{equation}\label{Uone}
J=\frac{i}{2} \left[e^{i\sqrt{2}\,x_L^1} + e^{-i\sqrt{2}\,x_L^1} + e^{i\sqrt{2}x_L^2} + e^{-i\,\sqrt{2}\,x_L^2}\right]\
\end{equation}
and a corresponding antiholomorphic current $\bar{J}\,$.

The chiral algebra of one $A_2$ minimal model can be written in terms of a free fermion (for general $k $ this would be a $\mathbb{Z}_k$ parafermion) and a free $U(1)$ boson\footnote{In \eqref{Tsca} and \eqref{MMsca} we have adopted different conventions for the right-moving currents. This is done to ensure that the twist fields of the orbifold and the observables of the Landau-Ginzburg model (to be introduced later) belong respectively to the $(a,c)$ and $(c,c)$ ring, as in the usual conventions. This is a consistent choice because the mirror map $Q\to-Q\,, G^{\pm}\to G^{\mp}$ is an automorphism of the superconformal algebra. }:
\begin{align}\label{MMsca}
&T_1= -\frac{1}{2}(\partial \phi_1 \partial \phi_1 + \psi_1\partial \psi_1)&&\bar{T}_1= -\frac{1}{2}(\bar{\partial} \phi_1 \bar{\partial} \phi_1 + \bar{\psi_1}\bar{\partial} \bar{\psi}_1)\nonumber\\
&G_1^{\pm}= \frac{1}{\sqrt{2}} \psi_1 e^{\pm i \sqrt{2}\phi_{1L}}&&\bar{G}_1^{\pm}= \frac{1}{\sqrt{2}} \bar{\psi}_1 e^{\mp i \sqrt{2}\phi_{1R}}\\
&J_1=\frac{i}{\sqrt{2}}\partial \phi_1&&\bar{J}_1=\frac{i}{\sqrt{2}}\bar{\partial} \phi_1\nonumber
\end{align}
The chiral algebra of the tensor product of two minimal models (restricting attention for the moment to the left-moving sector) will contain the currents $T=T_1+T_2\,$, $G^\pm=G_1^\pm + G_2^\pm$ and two $U(1)$ currents $ J_\pm=J_1\pm J_2$. 
One can check that the left-moving currents in  \eqref{Tsca} take this form under the identifications 
\begin{equation}
\begin{split}
&i\partial x^\pm=\frac{1}{\sqrt{2}} (\psi_1 e^{\mp iH_L} + \psi_2\,e^{\pm iH_L}) \qquad\text{with}\qquad H_L=\frac{1}{\sqrt{2}}(\phi_1-\phi_2)\\
&B_L=\frac{1}{\sqrt{2}}(\phi_1+\phi_2)\,,
\end{split}
\end{equation}
which map the $U(1)$ current \eqref{Uone} of the orbifold CFT to the current $J_-=i\partial H$ of the $A_2 \otimes A_2$ theory. 
The map for the right-moving currents differs by a sign in the expressions for $H_R$ and $B_R$,
\begin{align}
& H_R=-\frac{1}{\sqrt{2}}(\phi_1-\phi_2)
&B_R=-\frac{1}{\sqrt{2}}(\phi_1+\phi_2)\,,
\end{align}
because of the different conventions adopted for the right-moving currents in \eqref{Tsca} and \eqref{MMsca}. In order to prove that the models are equivalent one also needs to match the primary fields and their OPE's and this is done in \cite{Chun}.  In what follows we will look only at the spectrum of chiral primaries, since we are interested in comparing the topological observables on the two sides. More details about the complete map between the primary fields are collected in the appendix.

As a first step we need a characterization of the twisted sectors of the $T^2/\mathbb{Z}_4$ orbifold \cite{Dixon, Chun}\,. For the moment we restrict attention to the bosonic fields.
There are two $\mathbb{Z}_4$ fixed points, with coordinates $(0,0)$ and $(\pi R, \pi R)\,$. These correspond to the two conjugacy classes of the space group of translations and order one $\mathbb{Z}_4$ rotations. We denote the bosonic part of the twist fields associated with these two fixed points by $\sigma_0^{(\frac{1}{4})}$ and $\sigma_1^{(\frac{1}{4})}\,$. There are also four $\mathbb{Z}_2$ fixed points, corresponding to the conjugacy classes of the order two rotations (reflections) and translations. Correspondingly, we have four $\mathbb{Z}_2$ twist fields, denoted by $ \sigma_{00}^{(\frac{1}{2})}\,$, $ \sigma_{11}^{(\frac{1}{2})}\,$, $\sigma_{10}^{(\frac{1}{2})}\,$ and $ \sigma_{01}^{(\frac{1}{2})}$. The first two are associated with the $\mathbb{Z}_4$ fixed points at $(0,0)$ and $(\pi R, \pi R)\,$, while the other two are associated with the fixed points at $(0,\pi R)$ and $(\pi R, 0)\,$, which are exchanged by the $\mathbb{Z}_4$ action. The $\mathbb{Z}_4$ invariant twist fields are therefore $ \sigma_{0}^{(\frac{2}{4})}\equiv  \sigma_{00}^{(\frac{1}{2})}$, $\sigma_{1}^{(\frac{2}{4})}\equiv  \sigma_{11}^{(\frac{1}{2})}\, $ and $\sigma_{01}^{(\frac{2}{4})}\equiv \frac{1}{\sqrt{2}} (\sigma_{01}^{(\frac{1}{2})} +\sigma_{10}^{(\frac{1}{2})})\,$. 

\begin{figure}
\begin{center}
\includegraphics[scale=0.7]{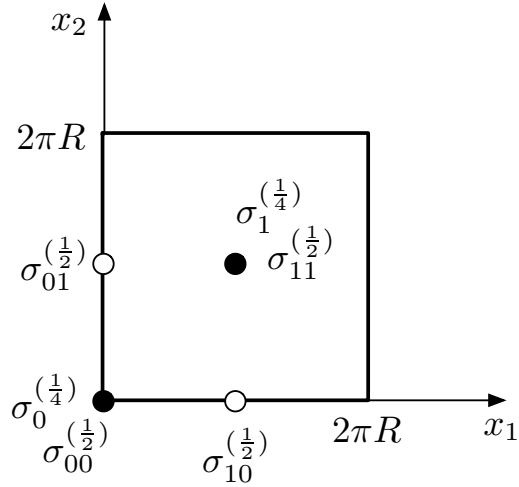}\caption{Fixed points and associated twist fields on $T^2/\mathbb{Z}_4\,$.}\label{twistfields}
\end{center}
\end{figure}

The twist fields of the full superconformal theory are obtained multiplying by $e^{ikB/4}$ the bosonic twist fields of the $k$-th twisted sector. A convenient basis of operators for the chiral ring (listed with the respective $(h,q;\bar{h},\bar{q})$ eigenvalues) is
\begin{equation}\label{acring}
\begin{array}{llllll}
\II &(0,0;0,0)&\quad \psi^{+}\bar{\psi}^{-}&\left(\frac{1}{2},1;\frac{1}{2},-1\right)\\ 
\Sigma_{\pm\frac{1}{4}}\phantom{\Big(}&\left(\frac{1}{8},\frac{1}{4}; \frac{1}{8},-\frac{1}{4}\right)&\quad\Sigma_{\pm\frac{2}{4}}, \Sigma_{0} &\left( \frac{1}{4},\frac{1}{2};\frac{1}{4},-\frac{1}{2}\right)&\quad\Sigma_{\pm\frac{3}{4}}& \left(\frac{3}{8},\frac{3}{4}; \frac{3}{8},-\frac{3}{4}\right)\ ,\end{array}
\end{equation}
where we have introduced linear combinations of twist fields\footnote{The linear combinations taken here differ from \cite{Chun} by a factor of $\sqrt{2}$ in the expression for $\Sigma_0\,$ (a possible misprint in the reference).} with definite charge with respect to the $U(1)$ current \eqref{Uone}:
\begin{align}\label{Sigmas}
&\Sigma_{+\frac{k}{4}}\equiv\frac{e^{ikB/4}}{\sqrt{2}} \left( \sigma_0^{(\frac{k}{4})}- i\, \sigma_1^{(\frac{k}{4})}\right)\qquad\qquad\Sigma_{-\frac{k}{4}}\equiv\frac{e^{ikB/4}}{\sqrt{2}} \left( -i\sigma_0^{(\frac{k}{4})}+\, \sigma_1^{(\frac{k}{4})}\right)\qquad\qquad k=1,3\nonumber
\\
&\Sigma_{\pm\frac{2}{4}}\equiv\frac{e^{iB/2}}{\sqrt{2}} \left(\pm\, \frac{i}{\sqrt{2}}\,\sigma_0^{(\frac{2}{4})}\mp \frac{i}{\sqrt{2}}\, \sigma_1^{(\frac{2}{4})} + \sigma_{01}^{(\frac{2}{4})}\right)\qquad \qquad\Sigma_{0}\equiv\frac{e^{iB/2}}{\sqrt{2}}\left(\sigma_0^{(\frac{2}{4})} +\phantom{i\,} \sigma_1^{(\frac{2}{4})}\right)\ .
\end{align}
All these chiral primaries are associated with K\"ahler deformations of the model:
the primary field $\psi^{+}\bar{\psi}^{-}$ is a marginal operator associated with the rescaling of the radius of the torus; all the other primaries are relevant operators, which from the spacetime point of view would be tachyons localized at the orbifold fixed points.  

These operators are in one-to-one correspondence with the chiral primaries of the $A_2\otimes A_2$ theory. 
In general, for an $A_k$ minimal model, the primary fields are constructed from the order parameters of the parafermionic system used in the free field representation of the model. In the simple case we are considering the parafermions are just free fermions and there is only one order parameter, the spin variable $\sigma$ of the Ising model, with $h=\bar{h}=\frac{1}{16}\,$. The chiral ring is generated by the identity and the operator $Y\equiv \sigma e^{i\frac{\phi}{2\sqrt{2}}}\,$, which has $h=\bar{h}= \frac{1}{8}$ and $q=\bar{q}=\frac{1}{4}\,$. One can check that $Y^2= e^{i\frac{\phi}{\sqrt{2}}}$ and $Y^3=0\,$, so the ring has the structure of $\frac{\mathbb{C}[Y]}{Y^3}\,$.
After tensoring two copies of the model we obtain the following list of chiral primaries:
\begin{equation}\label{ccring}
\begin{array}{llllll}
\II &(0,0:0,0)&\quad Y_1^2Y_2^2&\left(\frac{1}{2},1;\frac{1}{2},1\right)&&\\ 
Y_1, Y_2\phantom{\Big(}&\left(\frac{1}{8},\frac{1}{4};\frac{1}{8},\frac{1}{4}\right)&\quad Y_1^2,\, Y_2^2, \,Y_1Y_2 &\left( \frac{1}{4},\frac{1}{2};\frac{1}{4},\frac{1}{2}\right)&\quad Y_2^{\phantom{2}}Y_1^2,\, Y_1^{\phantom{2}}Y_2^2& \left(\frac{3}{8},\frac{3}{4};\frac{3}{8},\frac{3}{4}\right)\ .\end{array}
\end{equation}
This reproduces the list given above in \eqref{acring} for $T^2/\mathbb{Z}_4\,$, up to the sign of the right-moving $U(1)$ charge, so we have a map between the $(a,c)$ ting of the orbifold CFT and the $(c,c)$ ring on the minimal model side.

The $A_2 \otimes A_2$ model has a description as a Landau-Ginzburg model with superpotential $W_{LG}=Y_1^4+Y_2^4$ and the fields $Y_i$ in \eqref{ccring} can be identified with the lowest components of the LG superfields.  The chiral ring of the LG model is $\frac{\mathbb{C}[Y_1,Y_2]}{\partial_1W \partial_2 W}\,$, which is the same structure we see in \eqref{ccring}. As usual, mirror symmetry maps K\"ahler deformation on the A-side to polynomial deformations of the LG superpotential on the B-side. Turning on  the marginal deformation $W_{LG}= Y_1^4+Y_2^4 +2 \alpha Y_1^2Y_2^2\,$, which corresponds to deforming the $T^2/\mathbb{Z}_4$ orbifold away from the selfdual radius point, the relations that define the chiral ring of the LG model become 
\begin{equation}
Y_1^3 +\alpha Y_1Y_2^2=0\qquad \text{and}\qquad Y_2^3 +\alpha Y_2Y_1^2=0\,.
\end{equation}
This means that away from the self-dual radius point the mirror model cannot be written as the tensor product of two separate rational CFT's.

It is interesting to compare the symmetries of the two models. The LG model is invariant under exchange of the variables $Y_1$ and $Y_2$ and (for $\alpha=0$)  under an independent $\mathbb{Z}_4$ rotation of the two variables. These symmetries are also  present in the mirror model. The $Y_1\leftrightarrow Y_2$ symmetry is reflected in the CFT as invariance under exchange of the two fixed points associated with the twist fields $\sigma_0^{(\frac{k}{4})}$ and $\sigma_1^{(\frac{k}{4})}\,$. This is part of the quantum symmetry, together with a diagonal $\mathbb{Z}_4$ symmetry acting on the twist fields with weight equal to the order of the twist field:
\begin{align}
&\Sigma_i^{(\frac{k}{4})}\rightarrow e^{+2\pi i k/4}\, \Sigma_i^{(\frac{k}{4})}\qquad\qquad i=1,2\,.
\intertext{Due to the enhanced symmetry generated by the current \eqref{Uone}, the $\mathbb{Z}_4$ transformation}
&\Sigma_0^{(\frac{k}{4})}\rightarrow e^{+2\pi i k/4}\, \Sigma_0^{(\frac{k}{4})}\nonumber\\
&\Sigma_1^{(\frac{k}{4})}\rightarrow e^{-2\pi i k/4}\, \Sigma_1^{(\frac{k}{4})}\,
\end{align}
is also a symmetry. This accounts for the additional $\mathbb{Z}_4$ symmetry that we observe in 
the LG model, which at the conformal point is a subgroup of the symmetry generated by the $U(1)$ current $J_-$ of $A_2\otimes A_2\,$. As in \eqref{Sigmas}, the $T^2/\mathbb{Z}_4$  fields that map to the$A_2\otimes A_2$ (or Landau-Ginzburg) variables are those that have a definite $U(1)$ charge. 

We conclude this section with  a brief comment on the GSO projection. As explained in \cite{BGpermutation}, the B-branes that we will consider in the next section in the context of the Landau-Ginzburg model are consistent with the type 0A GSO projection in the $A_2\otimes A_2 $ CFT. Throughout the paper we will always work with two spacetime dimensions (from the $T^2/\mathbb{Z}_4$ point of view) and the boundary state that we will write in section \ref{MMBS} will be consistent with the type 0 projection. One can of course embed the model in a ten-dimensional string theory, but as it stands the orbifold we are considering will give rise to a nonsupersymmetric spacetime theory.

\section{B-branes in the Landau-Ginzburg model}\label{LGbranes}

In this section we look at the brane content of the model introduced in the previous section. Specifically, we are interested in constructing A-branes on $T^2/\mathbb{Z}_4\,$, which are expected to be in one-to-one correspondence with the B-branes in the mirror CFT. The problem of constructing and classifying these branes can be approached from at least three different angles: we can study directly the A-type boundary conditions on the $T^2/\mathbb{Z}_4$ orbifold, or we can consider the mirror theory $A_2\otimes A_2$ and describe the branes in this non-geometric background. Furthermore, the B-branes in minimal models can be described in terms of boundary states or, in the Landau-Ginzburg framework, in terms of matrix factorizations of the LG superpotential $W_{LG}\,$. In this section we start from this last point of view and later we will make contact with the other approaches. In particular, we will see in the next sections how to give a geometric interpretation (on $T^2/\mathbb{Z}_4$) to the LG branes constructed below.

\subsection{A brief review of matrix factorizations}

The topological B-branes of a  Landau-Ginzburg model are classified by the matrix factorizations of $W_{LG}\,$ \cite{K-LiAG, Brunner, K-Li}. A brane is completely characterized by a matrix
\begin{equation}\label{fact}
D=\left[\begin{array}{cc}0&F\\ G&0\end{array}\right]\quad \text{such that} \quad D^2=\left[\begin{array}{cc} F\cdot G &0\\ 0&G\cdot F \end{array}\right]=\left[\begin{array}{cc} W_{LG}\cdot\II_k &0\\ 0&W_{LG}\cdot\II_k \end{array}\right]\, ,
\end{equation}
where $F$ and $G$ are rank $k$ matrices with polynomial entries. This is a very abstract description, but it can be related to more usual constructions by interpreting the elements of the matrices $F$ and $G$ as boundary potentials in the LG model defined on a strip. Then one can show that the condition that $F$ and $G$ are a factorization of $W_{LG}$ is equivalent to requiring invariance of the action under B-type supersymmetry \cite{K-LiAG,Brunner,K-Li}. 

In this framework, the problem of finding the boundary chiral ring associated with any pair of B-type boundary conditions is translated into an algebraic problem, which only involves the explicit knowledge of the factorizations of $W_{LG}$ associated with the two boundaries.
The boundary observables in the chiral ring are a subset of the open string states of the physical theory and carry some relavant information, such as the number of moduli (to first order) of a given brane. Given the matrix $D\,$, one can also easily compute bulk and boundary correlators involving B-model topological observables, which compute certain protected quantities in the physical theory. For instance, the charges of the B-branes under the Ramond-Ramond ground states is computed in the topological LG framework by the residue formula \cite{K-Li}
\begin{equation}\label{disc}
\langle\mathcal{O}\rangle_{\text{disk}}= \frac{16}{2(2\pi i)^2}\oint \frac{\mathcal{O}\cdot \text{STr}[\partial_1 D\wedge \partial_2 D]}{\partial_1 W \partial_2 W}\,,
\end{equation} 
with $\mathcal{O}$ any of the B-model observables described in the previous section. Here we have written the formula for the case in which $W$ is a polynomial in two variables $Y_1$ and $Y_2\,$, which is what we will use, and the overall normalization has been fixed for convenience. 

Finally, it is important to mention that not all factorizations of the form \eqref{fact} correspond to different, or independent, branes. Two factorizations are equivalent if the corresponding matrices  $D$ and $D'$ can be related by an invertible linear transformation. Two factorizations are also equivalent if they are related by the exchange of $F$ and $G\,$: in this case they are interpreted as a brane - anti-brane pair. (The "anti-brane" here is a factorization that has boundary Witten index and charges, computed as in \eqref{disc}, equal to those of the original brane but of opposite sign). Moreover, it is possible to introduce in the LG formalism the notion of bound state, i.e. it is possible to give a prescription to construct, from two factorizations, a third one that describes the bound state of the two original branes. The goal is thus  to identify a minimal set of branes such that all other branes can be obtained from these as bound states (or anti-branes of bound states). 

This was just a very brief summary of the main features of the formalism that we will use. More details can be found, for example, in \cite{Brunner, K-Li}. 

\subsection{B-branes of $W_{LG}=Y_1^4+Y_2^4\,$}

We now apply to $W_{LG}=Y_1^4+Y_2^4\,$ the construction just outlined. The superpotential is of the form $W_{LG}=W_1(Y_1)+W_2(Y_2)\,$, which reflects the fact that the corresponding CFT is a tensor product of two minimal models. Note that since this is not a rational CFT, a full classification of boundary conditions is much harder to obtain than for a single minimal model. In this section we are not working in the boundary state formalism, but it is useful to keep track of how the various results translate between the two formalisms. 

The simplest boundary states that can be constructed in $A_2 \otimes  A_2$ are tensor products of boundary states of the two $A_2$ minimal models (this gives the so-called ``rational" boundary state). Such a construction also exists in the LG formalism: there is a prescription for writing a factorization of  $W_{LG}$ starting from two factorizations of the one variable superpotentials $W_1$ and $W_2\,$ \cite{Ashok}. We will call the branes obtained through this construction ``tensor product branes". In fact, all the branes in this class can be obtained as bound states of a single matrix factorization. This follows from the fact that for superpotentials of the form $W_{LG}=Y^n$ all B-branes can be generated from the linear factorization $F=Y\,$, $G=Y^{n-1} $ \cite{Orlov, HLL}. Therefore, since we are trying to identify a set of minimal branes, we only need to consider the tensor product of two such linear factorizations of $W_1$ and $W_2\,$. The corresponding factorization has rank two and is of the form 
\begin{equation}\label{tensfact}
F=\left[\begin{array}{cc}Y_1&Y_2\\ Y_2^3&-Y_1^3\end{array}\right] \qquad G=\left[\begin{array}{cc}Y_1^3&Y_2\\ Y_2^3&-Y_1\end{array}\right]\ .
\end{equation}
In the minimal model language this corresponds to the $|L=0\rangle\equiv |L_1=0\rangle\otimes |L_2=0\rangle$ boundary state (see the appendix of \cite{Brunner} for the notation).

Let us now discuss some properties of these branes, which can be derived in the framework of matrix factorizations.
Using the residue formula \eqref{disc} one can check that all the topological disc correlators without boundary insertions vanish. This means that the corresponding physical branes are not charged under the Ramond ground states and this is in fact confirmed from the boundary state analysis \cite{Brunner,K-LiMM}. From the point of view of the mirror model $T^2/\mathbb{Z}_4\,$,  we learn in particular  that the D1-branes that are mirror to the tensor product factorizations are not charged under Ramond twisted sector fields. We can also look at the spectrum of boundary operators, which can be computed as the cohomology of a BRST operator constructed from $D\,$. Leaving out the details of this computation, the result is that for the factorization \eqref{tensfact} the boundary Witten index vanishes. Moreover, one finds that the boundary chiral rings contains two fermionic modes, which are associated to moduli of the brane. Through mirror symmetry, each of them should be identified with the position and Wilson line of a D1-brane, so we can conjecture that the factorization \eqref{tensfact} describes in fact a superposition of two branes, which are free to move independently.  

\begin{table}
\begin{center}
\begin{tabular}{|l|c|c|c|}
\hline
\multicolumn{4}{|l|}{$\mathbf{F^{(1)}_k \equiv(Y_1-\eta_k Y_2)}\phantom{\Big(}$:}\\
\hline
$\eta_k\phantom{\Big(}$&\small{$q_{0}\equiv \frac{i}{2}\langle Y_1^2-Y_2^2-i\sqrt{2}Y_1Y_2 \rangle$}
&\small{$q_{1}\equiv \frac{i}{2}\langle -Y_1^2+Y_2^2-i\sqrt{2}Y_1Y_2 \rangle$}&\small{$q_{01}\equiv\frac{1}{\sqrt{2}}\langle Y_1^2+Y_2^2\rangle$}\\ \hline
$e^{i\pi/4}\phantom{\Big(}$&$\phantom{-}0$&$-i\sqrt{2}$&$-i$\\
$e^{-i\pi/4}\phantom{\Big(}$&$\phantom{-}i\sqrt{2}$&$\phantom{-}0$&$\phantom{-}i$\\
$e^{i3\pi/4}\phantom{\Big(}$&$\phantom{-}0$&$\phantom{-}i\sqrt{2}$&$-i$\\
$e^{-i3\pi/4}\phantom{\Big(}$&$-i\sqrt{2}$&$\phantom{-}0$&$\phantom{-}i$\\
\hline
\end{tabular}\\ \ \\
\begin{tabular}{|l|c|c|c|}
\hline
\multicolumn{4}{|l|}{$\mathbf{F^{(2)}_k\equiv (Y_1-e^{i\pi/4} Y_2) (Y_1-\eta_k Y_2)}\phantom{\Big(}$:}\\
\hline
$\eta_k\phantom{\Big(}$&\small{$q_{1}\equiv \frac{i}{2}\langle Y_1^2-Y_2^2-i\sqrt{2}Y_1Y_2 \rangle$}&\small{$q_{0} \equiv \frac{i}{2}\langle -Y_1^2+Y_2^2-i\sqrt{2}Y_1Y_2 \rangle$}&\small{$q_{01}\equiv\frac{1}{\sqrt{2}}\langle Y_1^2+Y_2^2\rangle$}\\\hline
$e^{-i\pi/4}\phantom{\Big(}$&$\phantom{-}i\sqrt{2}$&$-i\sqrt{2}$&$\phantom{\Big(}0\phantom{\Big(}$\\
$e^{3i\pi/4}\phantom{\Big(}$&$\phantom{-}0$&$\phantom{-}0$&$\phantom{\Big(}-2i\phantom{\Big(}$\\
$e^{-3i\pi/4}\phantom{\Big(}$&$-i\sqrt{2}$&$-i\sqrt{2}$&$\phantom{\Big(}0\phantom{\Big(}$\\
\hline
\end{tabular}\caption{Charges of the B-branes associated with the polynomial factorizations of $W_{LG}\,$. The notation is $q_i\equiv \langle e^{iB/2}\sigma_i^{(\frac{2}{4})}\rangle$ and the expressions in terms of the LG observables are obtained inverting  \eqref{Sigmas}. The results in this table can be interpreted geometrically as indicating which fixed points each branes passes through (see Figure \ref{branes}).  }\label{pippo}
\end{center}
\end{table}

The tensor product branes only account for a small subset of all the B-branes in the theory, but they are interesting because the tensor product construction makes it easy to derive their properties (see \cite{Ashok} for more details about the LG computations). Another interesting and easily tractable class of factorizations is given by polynomial (rank one) factorizations of the form
\begin{equation}\label{poly}
W_{LG}= \prod_{k}(Y_1 -\eta_k Y_2 ) \,,
\end{equation} 
where $\eta_k\equiv e^{ik\pi/4}\,$, $k\in \{\pm 1\,,\pm 3\}\,$, is a fourth root of $-1\,$. The properties of these branes were first discussed in \cite{Ashok} and the conformal field theory description, as permutation boundary states \cite{Recknagel},  was established in \cite{BGpermutation} (see also \cite{permutationII}). We will come back to the boundary state decription in section \ref{MMBS}. Since exchanging $F$ and $G$ only amounts to exchanging a brane with its anti-brane, from factorizations of this form we seem to obtain a total of seven independent B-branes. We introduce the notation $F^{(1)}_k\equiv (Y_1-\eta_k Y_2)$  for the linear factorizations and $F^{(2)}_k\equiv (Y_1-e^{i\pi/4} Y_2) (Y_1-\eta_k Y_2)\,$, with $\eta_k\neq e^{i\pi/4}\,$, for the  quadratic factorizations. In fact it was shown in \cite{BGpermutation} that the branes $F^{(2)}_k$ can be obtained as bound states of the linear factorizations $F^{(1)}_k\,$. Moreover, the same reference shows that the annihilation of a brane $F^{(1)}_k\,$ and its antibrane can produce a tensor product brane, if the tachyon that drives the annihilation has a  specific notrivial profile \footnote{We should point out that all these  statements are made in the context of the topological theory, so while they reflect correctly a relation between the branes of the physical theory, they are not statements about the dynamics. In order to know if a decay actually happens in the physical theory, more information is needed.}. It is not clear if there are more general branes that are not obtained from the linear polynomial factorizations, but it is possible that they constitute a complete set of minimal branes for this model. Some hints in this direction also come from the geometric interpretation of these branes, as we will see in the next sections. In any case, due to practical limitations, we will restrict attention to this class of branes.  

Unlike the tensor product branes, the branes described by the polynomial factorizations \eqref{poly} couple to some bulk observables. However, only the quadratic bulk operators $Y_1^2,Y_1Y_2,Y_2^2$ have a non-vanishing disc one-point function. From the previous section we know that these operators correspond to $\mathbb{Z}_2$ twist fields on $T^2/\mathbb{Z}_4\,$, so we learn that these branes carry some twisted sector Ramond charges, but they do not couple to the Ramond fields in the untwisted sector. The values of the non-zero one-pointfunctions are collected in Table \ref{pippo}.

It will also be useful to have some information about the spectrum of boundary operators. We skip the computation and just summarize the results. There are three even boundary preserving operators associated with each of the linear factorizations $F_k^{(1)}$ and four associated with each of the quadratic factorizations $F^{(2)}_k\,$. There are no odd boundary preserving operators, which means that these branes have no moduli. However, there is a single odd boundary changing operators for every pair of branes $F^{(1)}_k$ and $F^{(1)}_j\,$, $k\neq j\,$.  This information is summarized in the intersection matrices
\begin{align}\label{I1}
&I(F^{(1)}_k,F^{(1)}_j)=\left[\begin{array}{cccc}
\phantom{-}3&-1&-1&-1\\
-1&\phantom{-}3&-1&-1\\
-1&-1&\phantom{-}3&-1\\
-1&-1&-1&\phantom{-}3
\end{array}\;\right]
&I(F^{(2)}_k,F^{(2)}_i)=\left[\begin{array}{ccc}
4&0&0\\
0&4&0\\
0&0&4
\end{array}\right]\ . 
\end{align}
In addition, for $\eta_k=e^{i\pi/4}$ or $\eta_k=\eta_j\,$, the spectrum of physical open string states contains two even boundary operators that change $F^{(1)}_k$ into $F^{(2)}_j\,$.  For all the other values of $\eta_k$ and $\eta_j$ one finds instead two odd operators. A symmetric result holds for the open strings stretched in the opposite direction, from $F^{(2)}_j$ to $F^{(1)}_k\,$. The intersection matrices are
\begin{align}\label{I2}
&I(F^{(1)}_k,F^{(2)}_j)=\left[\begin{array}{cccc}
2&\phantom{-}2&-2&-2\\
2&-2&\phantom{-}2&-2\\
2&-2&-2&\phantom{-}2
\end{array}\right]
&I(F^{(2)}_j,F^{(1)}_k)=\left[\begin{array}{ccc}
\phantom{-}2&\phantom{-}2&\phantom{-}2\\
\phantom{-}2&-2&-2\\
-2&\phantom{-}2&-2\\
-2&-2&\phantom{-}2
\end{array}\;\right]\ .
\end{align}
One can verify that these intersection matrices are invariant under the full symmetry group of the LG model. 

\section{A-branes in the orbifold}\label{Abranes}

\begin{figure}
\begin{center}
\includegraphics[scale=0.6]{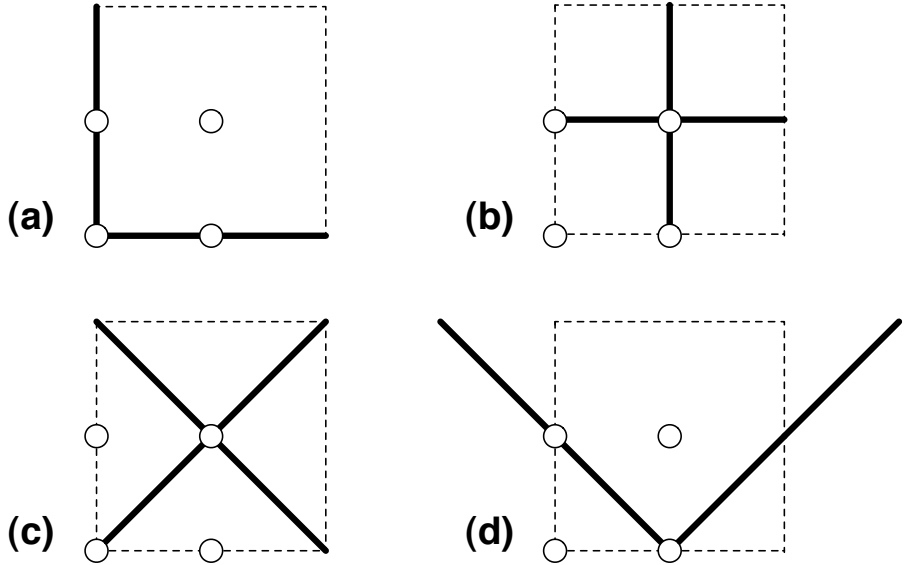}\caption{A-branes on $T^2/\mathbb{Z}_4\,$. This is a very schematic picture of the branes corresponding to the LG factorizations $F^{(1)}_k\,$, deduced from the values of the one-point function collected in Table \ref{pippo}. The dots denote the orbifold fixed points, as in Figure \ref{twistfields}. }\label{branes}
\end{center}
\end{figure}

In this section we take a different point of view and look at D-branes on $T^2/\mathbb{Z}_4\,$ that preserve A-type supersymmetry.
Ideally we would like to construct boundary states that reproduce the properties - in particular the couplings to the bulk fields - of the LG branes described in the previous section. We are especially interested in finding a geometric description of the LG branes associated with the polynomial factorizations \eqref{poly}. 
We found in the previous section that these branes are charged under some twisted sector fields, so we need to look in the orbifold CFT for a set of A-branes with this property. Geometrically, A-branes wrap middle-dimensional cycles, so we consider D1 branes that wrap one direction inside $T^2\,$, summing over images to implement the orbifold projection. For a generic cycle we would have to sum over four images, but we can find some special cases, as in Figure \ref{branes}, in which it is sufficient to sum over two images\footnote{In fact, if we keep into account the orientation of the branes, we still need to sum over four images, as will appear from the coefficient that measure the tension in the boundary states that we will derive in section \ref{MMBS}. However, for the purposes of this section it is useful to forget about the orientation, since it does not affect the computation of twisted sector couplings that we are interested in.}. The resulting brane should be rigid, because any displacement would lead to a configuration that doesn't respect the orbifold projection. This is consistent with the fact that the cycles in Figure \ref{branes} are precisely all those that pass through any two of the orbifold fixed points, so branes wrapping these cycles can carry charge under twisted sector fields. 

These observations suggest that the mirror of the B-branes corresponding to polynomial factorizations in the LG model could be, roughly,  A-branes wrapping the cycles of Figure \ref{branes}.
In the remaining part of the section we will try to give some arguments in favor of this conjectured identification. We will work in the boundary state framework, but we will not give a full boundary state description of the branes in Figure \ref{branes}. Rather, we will derive from spacetime considerations the topological couplings of these branes and show that they match the results of the previous section (we follow closely the general discussion  of reference \cite{OOY}). A more complete boundary state description of these branes will be given in the next section, exploiting the exact knowledge of the corresponding minimal model boundary states \cite{BGpermutation, permutationII}.  

Written in the closed string channel, the A-type boundary conditions are:
\begin{equation}\label{Abc}
\begin{split}
(T-\bar{T})|B\rangle=0\qquad\qquad
( G^+ +i\eta\, \bar{G}^-)|B\rangle=0 &\\
 ( J- \bar{J})|B\rangle=0\qquad\qquad
  (G^- +i\eta\, \bar{G}^+)|B\rangle=0& \,.
\end{split}
\end{equation}
We set $\eta=+1$ for consistency with the boundary condition $[Q_++\bar{Q}_+]_{\partial\Sigma}=0$ imposed in the LG model. 
If we are only interested in the topological couplings we can, following \cite{OOY,RS}, consider a linear combination of states that solve the conditions \eqref{Abc} in the NS sector:
\begin{equation}\label{Bca}
|B\rangle=\sum_ac^a \,\Bket{a}\,,
\end{equation}
where we only include the Ishibashi states corresponding to the allowed chiral primary fields. Here, for A-branes, the  label $a$ runs over the elements of the $(a,c)$ ring. The idea is that $|B\rangle$ will be related by spectral flow to the terms in the physical boundary state that encode the value of the topological couplings.

The relation between the coefficients $c^a$ in \eqref{Bca} and the disc one point functions of the A-model observables is found expanding \eqref{Bca}, neglecting all the terms involving descendants, and applying half a unit of spectral flow to implement the topological twist:
\begin{equation}\label{0top}
|B\rangle=\sum_ac^a \,|O_a\rangle + \ldots\,\xrightarrow{\text{A-twist}}\,\sum_ac^a\,\mathcal{O}_a\,|0\rangle_{top}\,.
\end{equation}
In the first equality we have pulled out of the boundary state the terms containing the chiral primaries, generically denoted $\mathcal{O}_a\,$, and in the final expression $|0\rangle_{top}$ is the topological vacuum, which coincides with a Ramond vacuum state of the untwisted theory. Therefore
\begin{equation}\label{ca}
c^a=\eta^{ab} \langle 0_{top}| \mathcal{O}_b |B\rangle\,,
\end{equation}   
where  $\mathcal{O}_b$ here stands for any A-model observable and  $\eta^{ab}$ denotes the inverse of the topological metric $\eta_{ab}\equiv \langle \mathcal{O}_a\mathcal{O}_b\rangle\,$ \cite{OOY}. 

In order to use the formula \eqref{ca} we will need to determine the metric $\eta_{ab}\,$.
It is easier to compute the two-point function on the B-side, in the LG model. Here the correlators of the bulk observables (in the theory without boundaries) are given by
\begin{equation}
\langle\mathcal{O}\rangle_{\text{sphere}} = \frac{16}{(2\pi i)^2 } \oint \frac{\mathcal{O}}{\partial_1W \partial_2W}\ ,
\end{equation} 
where again the unimportant overall factor has been fixed for convenience.
Applying this formula we find that the metric $\eta_{ab}$ only mixes the quadratic LG operators among themselves and on these subset of operators it looks like
\begin{equation}\label{eta}
\eta_{ab}=\left(\begin{array}{ccc}0&0&1\\0&1&0\\1&0&0\end{array}\right)\ , \qquad a,b=Y_1^2,\, Y_1Y_2,\, Y_2^2\,.
\end{equation}
From the CFT point of view, this is simply the statement that the $k$-th twisted sector only couples to the $(4-k)$-th twisted sector, with an additional constraint on the charges imposed by the enhanced $U(1)$ symmetry.

\subsection{Topological couplings}

Since the branes that we are trying to describe are only charged under Ramond fields in the second twisted sector (these are the observables $Y_1^2,Y_1Y_2, Y_2^2$ in LG language), for the purpose of deriving the topological couplings we can think of 
the boundary state \eqref{Bca} as being constructed in two steps. We can first write down some  boundary states invariant under the $\mathbb{Z}_2$ subgroup of the orbifold group and then, as a second step, superpose two such states to impose the full $\mathbb{Z}_4$ invariance. For concreteness let  us focus on a brane wrapping the cycle (a) in Figure \ref{branes}. We start by writing the boundary state for a brane entended in the $\hat{1}$  direction inside $T^2\,$, imposing for the moment only a $\mathbb{Z}_2$ projection. The conditions
\begin{align}\label{bc}
&(\partial x^1+\bar{\partial}x^1)|B\rangle =0&(\psi^1 + i \bar{\psi}^1)|B\rangle =0\nonumber\\
&(\partial x^2 -\bar{\partial}x^2)|B\rangle=0&(\psi^2 - i \bar{\psi}^2)|B\rangle=0 
\end{align} 
are solved in a  $\mathbb{Z}_2$ twisted sector by the coherent state
\begin{equation}
\Bket{\sigma^{(\frac{k}{2})}}_{\Hat{1}} = \exp\left[\sum_{r>0} \frac{1}{r} (-\alpha^1_{-r}\bar{\alpha}^1_{-r} + \alpha^2_{-r}\bar{\alpha}^2_{-r})+ i \sum_{r>0} (-\psi^1_{-r} \bar{\psi}^1_{-r}+\psi^2_{-r} \bar{\psi}^2_{-r})  \right] |{\sigma^{(\frac{k}{2})}}\rangle_{NS}\,,
\end{equation}
where $r\in \mathbb{Z}$ and $|{\sigma^{(\frac{k}{2})}}\rangle_{NS}$ denotes the NS vacuum in the sector twisted by $e^{ikB/2}{\sigma^{(\frac{k}{2})}}\,$, with the constraint $(\psi^1_0 + i \bar{\psi}_0^1)|{\sigma^{(\frac{k}{2})}}\rangle_{NS}=(\psi_0^2 - i \bar{\psi}_0^2)|{\sigma^{(\frac{k}{2})}}\rangle_{NS}=0$  from the zero modes boundary conditions. For the configuration we are considering, taking into account \eqref{ca} and \eqref{eta}, the relevant twisted sectors are those created by $e^{iB/2}\sigma^{(\frac{1}{2})}_{11}$  and $e^{iB/2}\sigma^{(\frac{1}{2})}_{01}\,$.
We thus write as an ansatz the linear combination of coherent states
\begin{equation}\label{ansBS}
\mathcal{N}\Big[ e^{i\theta_a}\,\Bket{\sigma^{(\frac{1}{2})}_{11}}_{\Hat{1}}  + \Bket{\sigma^{(\frac{1}{2})}_{01} }_{\Hat{1}} \Big] \,,
\end{equation}
with some normalization constant $\mathcal{N}\,$.

A complete analysis of boundary states for D1 branes extended in $S^1/\mathbb{Z}_2$ orbifolds was carried out in \cite{Sen}. In this section we do not aim to write the full boundary states for the branes on $T^2/\mathbb{Z}_4\,$, but we can borrow an argument given in \cite{Sen} to associate a physical meaning to the phase that appears in \eqref{ansBS}. This in turn will allow us to give an interpretation to the topological one-point functions computed in the previous section. The argument can be summarized as follows. The two fixed points of the $S^1/\mathbb{Z}_2$ orbifold correspond to the two conjugacy classes of the group $\Lambda \ltimes \mathbb{Z}_2$ where $\Lambda $ denotes the group of translations by $2\pi R\,$. If we consider the transformations
\begin{equation}
\begin{split}
&g_{0}: \quad x \,\to\, -x\\  &g_{1}: \quad x\,\to\, 2\pi R-x\,,
\end{split}
\end{equation}
such that
\begin{equation}
g_{1}g_{0}:\quad x\, \to\, x+2\pi R\,,
\end{equation}
then we can think of  $\sigma^{(\frac{1}{2})}_{01}$ and $\sigma^{(\frac{1}{2})}_{11}$ as being associated with the sectors twisted by $g_{0}$ and $g_{1}$ respectively. In overlap computations, the terms coming from the overlap of the Ishibashi states $\Bket{\sigma^{(\frac{1}{2})}_{01}}$ are interpreted in the open string channel as terms coming from a trace with a $g_{0}$ insertion. Similarly, the overlap of   $\Bket{\sigma^{(\frac{1}{2})}_{11}}$ is associated with a term in the open string trace with a $g_{1}$ insertion. If we take a string stretched between two branes with Wilson lines $\theta$ and $\theta'\,$, the wave function of the open string states picks up a phase $e^{i(\theta-\theta')}$ under a translation by $2\pi R\,$. Under the action of $\mathbb{Z}_2$ the winding quantum number $w$ picks up a sign, so invariance of the unwtisted sector coherent state
\begin{equation}
\sum_{w} e^{i\theta w} \Bket{w} \,
\end{equation} 
restricts the choice of Wilson line to the values $\theta=0,\pi\,$.  Therefore, from what we have said so far, for a string stretched between two branes with the same $\theta$ the action of  $g_{0}$ and $g_{1}$ should be identical, and so the terms containing $\Bket{\sigma^{(\frac{1}{2})}_{01}}$  and $\Bket{\sigma^{(\frac{1}{2})}_{11}}$ should contribute equally to the overlap. Instead, if the value of $\theta$ is different for the two branes, then the contributions of the overlaps should differ by a sign, to account for the phase $e^{i(\theta-\theta')}\,$. 
In conclusion, as the notation suggests, in \eqref{ansBS} the parameter $\theta_a$ should be interpreted as the Wilson line on the brane, with $\theta_a=0,\pi\,$.

To impose the full $\mathbb{Z}_4$ projection we need to add the image of the brane under the transformation $x^1\to x^2, x^2\to -x^1\,$.  
For the image the discussion goes as before, so we can consider a linear combination of states
\begin{equation}\label{ansBS2}
\mathcal{N}\Big[e^{i\theta_a} \,\Bket{\sigma^{(\frac{1}{2})}_{11}}_{\Hat{2}} +\Bket{\sigma^{(\frac{1}{2})}_{10}}_{\Hat{2}}  \Big]\,.
\end{equation}
Note that the value of the Wilson line is again $\theta_a$ to respect the $\mathbb{Z}_4$ symmetry. 
Superposing \eqref{ansBS} and \eqref{ansBS2}, we conclude that the boundary state that describes a brane wrapping the cycle (a) in  Figure \ref{branes} must contain the Ramond sector terms related by spectral flow to   
\begin{equation}\label{BS1}
\mathcal{N}\Big[ e^{i\theta_{a}}\,\Bket{\sigma^{(\frac{1}{2})}_{11}}_{\Hat{1}} +e^{i\theta_{a}}\,\Bket{\sigma^{(\frac{1}{2})}_{11}}_{\Hat{2}}  +\Bket{\sigma^{(\frac{1}{2})}_{01}}_{\Hat{1}}  +  \Bket{\sigma^{(\frac{1}{2})}_{10}}_{\Hat{2}}   \Big] = \sqrt{2}\,\mathcal{N}\Big[\sqrt{2} e^{i\theta_{a}}\,|\sigma^{(\frac{2}{4})}_{11}\rangle + \, |\sigma^{(\frac{2}{4})}_{01}\rangle \Big]+\dots\,,
\end{equation}
where $\sigma^{(\frac{2}{4})}_{01}\equiv (\sigma^{(\frac{1}{2})}_{10}+ \sigma^{(\frac{1}{2})}_{01})/\sqrt{2}$ is the $\mathbb{Z}_4$ invariant combination of twist fields and we have expanded the coherent states as in \eqref{0top}. What we have gained by the analysis of the previous paragraph is that we could relate the coefficients in \eqref{BS1} to the value of the Wilson lines on the branes.

Using \eqref{ca}, we can extract from the expression \eqref{BS1} the ratio of couplings
$$
\frac{q_{0}}{q_{01}} = \sqrt{2} \,e^{+i\theta_{a}}\,
$$ 
and comparison with Table \ref{pippo} shows that this is the correct answer for the LG branes described by linear factorizations with $\eta= e^{-i\pi/4}, e^{-i3\pi/4}$ for the different choices of Wilson line $\theta_{a}\,$. This is consistent with the identification
\begin{subequations}\label{amap}
\begin{align}
&F^{(1)}_{-1}\equiv(Y_1-e^{-i\pi/4}\,Y_2)\qquad&&\longleftrightarrow&&\qquad |a,0\,\rangle\equiv +\sqrt{2} \,\Bket{\sigma^{(\frac{2}{4})}_{11}}  + \, \Bket{\sigma^{(\frac{2}{4})}_{01} } \phantom{\,.}\\
&F^{(1)}_{-3}\equiv(Y_1-e^{-i3\pi/4}\,Y_2)\qquad&&\longleftrightarrow&&\qquad |a,\pi\rangle\equiv -\sqrt{2} \,\Bket{\sigma^{(\frac{2}{4})}_{11} }+ \, \Bket{\sigma^{(\frac{2}{4})}_{01}} \,,
\end{align}
\end{subequations}
where now $\Bket{\sigma^{(\frac{2}{4})}_{11}}$ and $\Bket{\sigma^{(\frac{2}{4})}_{01}}$ are Ishibashi states of the $\mathbb{Z}_4$ orbifold. For simplicity here we have dropped the normalization factor.

Along the same lines, the configuration (b) in Figure \ref{branes} can be associated with the linear combination of coherent states
\begin{equation}\label{BS2}
\sqrt{2}\,\mathcal{N}\Big[\sqrt{2}\,e^{i\theta_{b}} \Bket{\sigma^{(\frac{2}{4})}_{00}}- \Bket{\sigma^{(\frac{2}{4})}_{01}}  \Big]\,,
\end{equation}
which is consistent with the map
\begin{subequations}\label{bmap}
\begin{align}
&F^{(1)}_{+1}\qquad\longleftrightarrow\qquad |b,\pi\rangle\equiv +\sqrt{2}\, \Bket{\sigma^{(\frac{2}{4})}_{00}} - \Bket{\sigma^{(\frac{2}{4})}_{01}}\phantom{\ .}\\
&F^{(1)}_{+3}\qquad\longleftrightarrow\qquad |b,0\,\rangle\equiv -\sqrt{2}\, \Bket{\sigma^{(\frac{2}{4})}_{00} } - \Bket{\sigma^{(\frac{2}{4})}_{01}}\ .
\end{align}
\end{subequations}
The sign in front of $\Bket{\sigma^{(\frac{2}{4})}_{01}}$ is required, by an argument similar (T-dual) to the one given above, because a phase $e^{i(x-x')/R}$ appears in the wave functions associated with open strings that stretch between the branes in (a) and (b).  

The configurations (c) and (d) in Figure \ref{branes} are related to the quadratic factorizations of the LG superpotential $W_{LG}\,$. The notation is $F^{(2)}_k\equiv (Y_1-e^{i\pi/4} Y_2) (Y_1-\eta_k Y_2)\,$.
For the configuration (c) we can write
\begin{equation}\label{BS3}
|c,\theta_{c}\rangle\equiv 2\,\mathcal{N}\Big[ \Bket{\sigma^{(\frac{2}{4})}_{00}}  - e^{i\theta_{c}}\Bket{\sigma^{(\frac{2}{4})}_{11}}  \Big]\,,
\end{equation}
which corresponds to the factorizations $F^{(2)}_{-1}$ and $F^{(2)}_{-3}$ for the two different choices of Wilson lines $\theta_{c}=0$ and $\theta_{c}=\pi\,$:
\begin{subequations}\label{cmap}
\begin{align}
&F^{(2)}_{-1}\qquad\longleftrightarrow\qquad |c,0\,\rangle\equiv  +\sqrt{2}\,\Bket{\sigma^{(\frac{2}{4})}_{00}}  - \sqrt{2}\, \Bket{\sigma^{(\frac{2}{4})}_{11}}  \phantom{\ .}\\
&F^{(2)}_{-3}\qquad\longleftrightarrow\qquad |c,\pi\rangle\equiv +\sqrt{2}\,\Bket{\sigma^{(\frac{2}{4})}_{00}} +\sqrt{2}\, \Bket{\sigma^{(\frac{2}{4})}_{11}} \ .
\end{align}
\end{subequations}
The remaining factorization $F^{(2)}_{-1}$ will thus correspond to configuration (d). In this case we don't have a Wilson line degree of freedom, since the $\mathbb{Z}_4$ symmetry requires the contribution from the two $\mathbb{Z}_2$ fixed points to be the same. Repeating the steps illustrated before, we can associate to  one ``component" of the brane the linear combination of Ishibashi states
\begin{align}
\mathcal{N}\Big[ \Bket{\sigma^{(\frac{1}{2})}_{10}}_{\Hat{1}+\Hat{2}} + \Bket{\sigma^{(\frac{1}{2})}_{01}}_{\Hat{1}+\Hat{2}} \Big] =\sqrt{2}\, \mathcal{N} \, \Bket{\sigma^{(\frac{2}{4})}_{01}}_{\Hat{1}+\Hat{2}}  \nonumber \,.
\end{align}
and the same for the image. As a result we find
\begin{equation}\label{dmap}
F^{(2)}_{+3}\qquad\longleftrightarrow\qquad|d\rangle\equiv 2\sqrt{2}\,\mathcal{N}\,\Bket{\sigma^{(\frac{2}{4})}_{01}}\ .
\end{equation}
Note that the normalization is consistent with the relative factor of $2$ that we observe in Table \ref{pippo} between the charge of this brane and the $q_{01}$ charge of the other branes. 

\subsection{Intersection matrix}

As a further check, we now compare the intersection matrix of the LG branes computed in section \ref{LGbranes} with the index $I(J,K)\equiv\text{Tr}_{J,K}(-1)^F e^{-\beta H}$ of the boundary CFT. In the closed string channel this is expressed as an overlap $I(J,K)=\phantom{,}_{R}\langle J|(-1)^{q_L}|K\rangle_{R}\,$, where $q_L$ is the left $U(1)$ R-chrage and the subscript $R$ indicates that we only take the Ramond terms (ground states) in the boundary state, properly normalized. Note that the ground states that contribute in our case will have $q_L=0$ (see Table \ref{RRgs} in the appendix). 

We begin by trying to reproduce the intersection matrix  
$I(F^{(1)}_k,F^{(1)}_j)\,$, which we found earlier to have the form \eqref{I1}. We find the matrix elements 
\begin{align}
\langle a,\theta'_{a}|a,\theta_{a}\rangle= 2 \,e^{i(\theta_{a}-\theta'_{a})}\Bovlap{\sigma^{(\frac{2}{4})}_{1}}{\sigma^{(\frac{2}{4})}_{1}}+\Bovlap{\sigma^{(\frac{2}{4})}_{01}}{\sigma^{(\frac{2}{4})}_{01}}=\quad &+3  \qquad \text{if}\qquad \theta_{a}=\theta_{a}'\nonumber\\
&-1\qquad\text{if}\qquad \theta_{a}\neq\theta_{a}'\nonumber\\
\langle b,\theta'_{b}|b,\theta_{b}\rangle= 2 \,e^{i(\theta_{b}-\theta'_{b})}\Bovlap{\sigma^{(\frac{2}{4})}_{0}}{\sigma^{(\frac{2}{4})}_{0}}+\Bovlap{\sigma^{(\frac{2}{4})}_{01}}{\sigma^{(\frac{2}{4})}_{01}}= \quad&+3 \qquad \text{if}\qquad \theta_{b}=\theta_{b}'\\
&-1\qquad\text{if}\qquad \theta_{b}\neq\theta_{b}'\nonumber\\
\langle a,\theta_{a}|b,\theta_{b}\rangle=\langle b,\theta_{b}|a,\theta_{a}\rangle=-\Bovlap{\sigma^{(\frac{2}{4})}_{01}}{\sigma^{(\frac{2}{4})}_{01}}= \quad&-1 \nonumber\,.
\end{align} 
One can check that these results agree with \eqref{I1} precisely if the identifications \eqref{amap} and \eqref{bmap} hold. Similarly, we find for
$I(F^{(2)}_k,F^{(2)}_j)$ the matrix elements
\begin{align}
\langle c,\theta'_{c}|c,\theta_{c}\rangle= 2\,e^{i(\theta_{c}-\theta'_{c})} \Bovlap{\sigma^{(\frac{2}{4})}_{1}}{\sigma^{(\frac{2}{4})}_{1}}+2\,\Bovlap{\sigma^{(\frac{2}{4})}_{0}}{\sigma^{(\frac{2}{4})}_{0}}=\quad +&4 \qquad \text{if}\qquad \theta_{c}=\theta_{c}'\nonumber\\&0\qquad\text{if}\qquad \theta_{c}\neq\theta_{c}'\nonumber\\
\langle d|d\rangle= +4 \Bovlap{\sigma^{(\frac{2}{4})}_{01}}{\sigma^{(\frac{2}{4})}_{01}}=\quad +&4 \\
\langle c,\theta_{c}|d\rangle= \langle d|c,\theta_{c}\rangle=\quad\phantom{+}&0\nonumber\,.
\,
\end{align} 
Using the identifications \eqref{cmap} and \eqref{dmap} these are seen to agree  with the answer  \eqref{I1} found from the LG computation. For completeness, we also write the result of the computation of $I(F^{(2)}_k,F^{(1)}_j)=I(F^{(1)}_j,F^{(2)}_k)\,$:
\begin{align}
\langle c,\theta_{c}|a,\theta_{a}\rangle= -2\,e^{i(\theta_{a}-\theta_{c})} \Bovlap{\sigma^{(\frac{2}{4})}_{1}}{\sigma^{(\frac{2}{4})}_{1}}=\quad -&2 \qquad \text{if}\qquad \theta_{c}=\theta_{a}\nonumber\\+&2\qquad\text{if}\qquad \theta_{c}\neq\theta_{a}\nonumber\\
\langle c,\theta_{c}|b,\theta_{b}\rangle= -2\,e^{i\theta_{b}}\Bovlap{\sigma^{(\frac{2}{4})}_{0}}{\sigma^{(\frac{2}{4})}_{0}}=\quad +&2 \qquad \text{if}\qquad \theta_{b}=\pi\\-&2\qquad\text{if}\qquad \theta_{b}=0\nonumber\\
\langle d|a,\theta_{a}\rangle=-2\, \Bovlap{\sigma^{(\frac{2}{4})}_{01}}{\sigma^{(\frac{2}{4})}_{01}}= \quad-&2\nonumber\\
\langle d|b,\theta_{b}\rangle=+2\, \Bovlap{\sigma^{(\frac{2}{4})}_{01}}{\sigma^{(\frac{2}{4})}_{01}}= \quad+&2\nonumber
\,,
\end{align} 
which is in agreement with \eqref{I2}. Note that here the intersection matrix appears to be symmetric by construction, whereas in the LG formalism this property is not manifest.

\section{Comparison with boundary states in $A_2\otimes A_2$}\label{MMBS}
 
So far we have looked at the properties of branes in the topological sector of $T^2/\mathbb{Z}_4$ and the mirror LG model $W_{LG}=Y_1^4+Y_2^4\,$, showing that they agree. In the topological sector this LG model is completely equivalent to the $A_2\otimes A_2$ CFT and we have mentioned earlier that the map between the LG B-branes and the boundary states in the CFT is known explicitly. More precisely, the statement is that one can construct some consistent boundary states, with B-type boundary conditions, that reproduce all the properties of the topological B-branes \cite{K-LiMM , Brunner,BGpermutation,permutationII}. However, the boundary states contain more information, since they are constructed in the physical theory. In this section we use the explicit knowledge of boundary states in the minimal models (and how they relate to the LG factorizations) to write down the boundary states for the A-branes on $T^2/\mathbb{Z}_4\,$. The goal is to complete and put on firmer ground the geometric picture of section \ref{Abranes}. 

\subsection{Permutation boundary states}\label{MMPB}

It was shown in \cite{BGpermutation} that the LG branes obtained from polynomial factorizations of the form \eqref{poly} are described in CFT language by permutation boundary states \cite{Recknagel}. In the case of the tensor product of two minimal models, these satisfy the B-type boundary conditions
\begin{align}\label{permbc}
&(T_1-\bar{T}_2)|B\rangle=0 &&(T_2-\bar{T}_1)|B\rangle=0\nonumber\\
&( G_1^\pm +i\eta\, \bar{G}_2^\pm)|B\rangle=0&&( G_2^\pm +i\eta\, \bar{G}_1^\pm)|B\rangle=0\\
& ( J_1+ \bar{J}_2)|B\rangle=0&&
 ( J_2+ \bar{J}_1)|B\rangle=0  \nonumber \ ,
\end{align}
which mix the currents of the two minimal models. Again, we take $\eta=+1$ for consistency with the combination of supercharges preserved by the LG branes. Through the map discussed in section \ref{CFT}, these boundary conditions can be rewritten in terms of $T^2/\mathbb{Z}_4$ currents. One finds simply
\begin{align}\label{permbcT}
&(T-\bar{T})|B\rangle=0 &&
( G^\pm +i\eta\, \bar{G}^\pm)|B\rangle=0\\
& ( Q- \bar{Q})|B\rangle=0&&
 ( J+ \bar{J})|B\rangle=0  \nonumber \ ,
\end{align}
where $Q$ is the $U(1)$ R-current and $J$ is the other $U(1)$ current, associated with the enhanced symmetry at the self-dual radius. These are, as expected, A-type boundary conditions. The choice of gluing with a nontrivial permutation in \eqref{permbc} is reflected here in the boundary condition imposed on $J\,$. As is familiar from the case of an unorbifolded compactification on a circle at the self-dual radius, where the geometry is effectively an $S^3\,$, here we can think of the geometry as containing an extra circle, associated with the current $J\equiv i\partial H\,$. The permutation and tensor product branes are distinguished by having, respectively, Dirichlet and Neumann boundary conditions in this direction.

The permutation boundary states \cite{Recknagel, BGpermutation, permutationII} are of the form 
\begin{equation}\label{pbs}
|[L,M,S_1,S_2]\rangle = \sum_{l,m,s_1,s_2} C^{L,M,S_1,S_2}_{l,m,s_1,s_2}\,\Bket{[l,m,s_1]\otimes [l,-m, -s_2]}^{\sigma} \,,
\end{equation}
where the superscript $\sigma$ distinguishes the Ishibashi states that solve the gluing conditions above and the coefficients are
\begin{equation}\label{pbsC}
C^{L,M,S_1,S_2}_{l,m,s_1,s_2}= \frac{1}{2\sqrt{2}}\, e^{i\pi Mm/4} e^{-i\pi(S_1s_1-S_2s_2)/2} \,\frac{\sin[\frac{\pi}{4} (L+1)(l+1)]}{\sin[\frac{\pi}{4}(l+1)]}\,.
\end{equation}
The quantum numbers $(l,m,s)$ that label the Ishibashi states take the values $l=0,1,2\,$, $m=-4, \ldots, +3\,$ (mod $8$), $s_i=0,\pm1,2\,$ (mod $4$), with the conditions that $l+m+s_1 $ and $s_1-s_2$ must be even. Note that in \eqref{pbs} it has been assumed that $l_1=l_2\equiv l$ and $m_1=-m_2\equiv m$, because this is necessary for the boundary conditions \eqref{permbc} to make sense. The square brackets in \eqref{pbs} express the fact that the quantum numbers are defined up to the equivalence relations 
\begin{align}
&(l,m,s_1) \sim (2-l,m+4,s_1+2)\nonumber\\
&(l,m,-s_2) \sim (2-l,m+4,-s_2+2)\label{equivtwo}\,.
\end{align}
These relations can be used to restrict the sum in \eqref{pbs} to $s_{1}=0,1\,$, while $s_2$ is still summed over all values $s_2=0,\pm1,2$ because $l_2, m_2$ are not independent labels in the construction of permutation Ishibashi states. The quantum numbers $[L,M,S_1,S_2]$ that label the boundary states take values in the same range as $[l,m,s_1,s_2]\,$, with an analogous equivalence relation. We set $S_1=S_2=0\,$, because this choice corresponds to the boundary conditions with $\eta=+1$ which have been implicitly used in the previous sections. With this choice, $L+M$ must be even.

\subsubsection{Ramond charges}

The value of the topological one-point functions computed in the LG formalism can be read off from the coefficients in the terms of the boundary state that correspond to Ramond ground states. These are the Ishibashi states in \eqref{pbs} with quantum numbers $s_1=s_2=1$ and $m=l+1\,$. For these, adjusting appropriately the normalization, from \eqref{pbsC} we have 
\begin{equation}\label{pbsTOP}
|[L,M,0,0]\rangle_{top} = \frac{1}{\sqrt{2}}\sum_{l=0,1,2} e^{i\frac{\pi}{4} M (l+1)}\, \frac{\sin\big[\frac{\pi}{4}\,(L+1)(l+1)\big]}{\sin\big[\frac{\pi}{4} (l+1)\big]}\,\Bket{[l,l+1,+1]\otimes [l,-l-1, -1]}^{\sigma} \!.
\end{equation}

In order to be able to compare with the results obtained previously from matrix factorizations, we need to know what are the LG observables that correspond to the quantum numbers that appear in these Ishibashi states and we need to know what is the map between LG factorizations and permutation boundary states. The latter was worked out in \cite{BGpermutation}, matching the intersection matrix obtained from the overlap of the boundary states with the LG results. The linear factorizations $F=Y_1-\eta Y_2$ are mapped to the boundary states with $L=0$ and $\eta= e^{-i\pi (M+1)/4}\,$. Since $L+M$ must be even, the label $M$ has only four independent values $M=-4,-2,0,+2\,$, so we find correctly that $\eta$ equals a fourth root of unity. In the notation of section \ref{LGbranes}, the map for the linear factorizations reads
\begin{equation}\label{mapkM}
F^{(1)}_{k}\qquad \longleftrightarrow\qquad |[0,M=-(k+1),0,0]\rangle\ .
\end{equation}
The permutation boundary states with $L=1$ are mapped to quadratic factorizations in the LG model; note however that not all quadratic factorizations seem to be realized in this way \cite{BGpermutation}. Finally, $L=2$ corresponds to the anti-branes of the linear factorizations. 

\begin{table}
\begin{center}
\begin{tabular}{|l|c|c|c|}\hline
\phantom{\Big(}&$\langle Y_1^2\rangle$
&$\langle Y_1Y_2\rangle$&$\langle Y_2^2\rangle$\\\hline
$F^{(1)}_{+1}\quad (M=-2)\phantom{\Big(}$ &$+\frac{1}{2}-\frac{i}{2}$&$-\frac{i}{\sqrt{2}}$&$-\frac{1}{2}-\frac{i}{2}$\\
$F^{(1)}_{-1}\quad (M=\phantom{-}0)\phantom{\Big(}$ &$+\frac{1}{2}+\frac{i}{2}$&$+\frac{i}{\sqrt{2}}$&$-\frac{1}{2}+\frac{i}{2}$\\
$F^{(1)}_{+3}\quad (M=-4)\phantom{\Big(}$ &$-\frac{1}{2}-\frac{i}{2}$&$+\frac{i}{\sqrt{2}}$&$+\frac{1}{2}-\frac{i}{2}$\\
$F^{(1)}_{-3}\quad (M=\phantom{-}2)\phantom{\Big(}$ &$-\frac{1}{2}+\frac{i}{2}$&$-\frac{i}{\sqrt{2}}$&$+\frac{1}{2}+\frac{i}{2}$\\\hline
\end{tabular}\caption{Charges of topological B-branes from permutation boundary states}\label{pluto}
\end{center}
\end{table}

Next, let us consider the meaning of the label $l$ in \eqref{pbsTOP}. As it was mentioned above, the Ramond ground states are in one-to-one correspondence with the chiral primaries of the model in the NS sector. For each of the two minimal models in $A_2 \otimes {A}_2$, these are the primaries with quantum numbers  $(l,m=l+1,s=1)\,$. The corresponding dimension and $U(1)$ charge are
\begin{equation*}
\begin{split}
&h_l=\frac{l(l+2)-l^2}{16}\,,\qquad\qquad q_l=\frac{l}{4}\,,\qquad\qquad l=0,1,2\,,
\end{split}
\end{equation*}
so that $h=\frac{q}{2}\,$. We are considering $(c,c)$ states, so this relation holds with the same sign for both left- and right-movers. As we already discussed in section \ref{CFT}, the chiral primaries that carry these quantum numbers are of the form $\big(\sigma e^{i\frac{\phi}{2\sqrt{2}}}\big)^l\sim Y^l\,$, where $\sigma$ and $\phi$ are the minimal model fields and $Y$ is the field that appears in the Landau-Ginzburg model.  The quantum numbers that appear in \eqref{pbsTOP} are schematically, writing explicitly both left- and right-moving sectors,
\begin{equation*}
(l,l+1,1)_1\otimes (l,-l-1,-1)_2 \otimes (l,l+1,1)_{\overline{1}} \otimes (l,-l-1,-1)_{\overline{2}}\ 
\end{equation*}
and using the equivalence relation \eqref{equivtwo} we can rewrite
\begin{equation*}
(l,-l-1,-1)_2\, \sim\, (2-l, -l-1+4, 1)_2\,.
\end{equation*}
Hence, in \eqref{pbsTOP}, the quantum number $l$ labels a subset of the chiral primaries of ${A}_2 \otimes {A}_2\,$, which are identified in LG language with the quadratic observables $Y_1^2,Y_1Y_2,Y_2^2 \,$. Specifically, we can take (see Table \ref{oneMM} and Table \ref{MMxMM} in appendix \ref{map})
\begin{equation}
l=0 \leftrightarrow Y_2^2\,,\qquad l=1 \leftrightarrow Y_1Y_2\,,\qquad l=2 \leftrightarrow Y_1^2\,.
\end{equation}
There is still room for some numerical factors, which are crucial in comparing the one-point functions. It seems to be necessary for consistency with the LG results to postulate a relation of the kind\footnote{I am very thankful to Ilka Brunner for her  help on this point.}
$$
\Bket{[l,l+1,1]_1\otimes [l,-l-1,-1]_2}= e^{i\pi (l+1)/4}\Bket{Y_1^{l}Y_2^{2-l}}\,.
$$ 
Putting all this together, we are able to reproduce the topological one-point functions computed in section \ref{LGbranes} from the boundary states \eqref{pbs}. Recalling the discussion around eq. \eqref{0top}-\eqref{eta}, we have
\begin{equation}
\langle\, Y_1^{2-l}Y_2^l \,\rangle_{L=0,M}=    \frac{e^{i\frac{\pi}{4} (M +1)(l+1)}}{\sqrt{2}}\ .
\end{equation}
The results are shown in table \ref{pluto} and are in agreement with the earlier results of section \ref{LGbranes}.

\subsubsection{Complete $T^2/\mathbb{Z}_4$ boundary states}

The main point of this section is that we can use the expression for the permutation boundary states in \eqref{pbs} to write down the complete boundary states (as opposed to having only the ``topological terms", as in the previous section) that describe the A-branes on $T^2/\mathbb{Z}_4\,$. This can be done because we have an explicit dictionary between the primary fields in the two models, which is summarized in the appendix. In particular, Table \ref{MMtoT} tells us how to associate to each Ishibashi state that appears in \eqref{pbs} a corresponding Ishibashi state of the $T^2/\mathbb{Z}_4$ CFT. 

In the NS sector we find that the Ishibashi states that appear in the boundary state \eqref{pbs} are:
\begin{align}\label{NSIsh}
&\Bket{[0,0,0]\otimes [0,0,0]}^\sigma &&\leftrightarrow&& \Bket{0,0;0,0}_{NS}\nonumber\\
&\Bket{[1,-1,0]\otimes [1,1,0]}^\sigma  &&\leftrightarrow&&\frac{e^{+i\pi/4}}{\sqrt{2}}\,\Big[\Bket{1,0;0,0}_{NS}^{\mathbb{Z}_4}-i\Bket{0,1;0,0}_{NS}^{\mathbb{Z}_4}\Big]\\
&\Bket{[1,1,0]\otimes [1,-1,0]}^\sigma &&\leftrightarrow &&\frac{e^{-i\pi/4}}{\sqrt{2}}\,\Big[\Bket{1,0;0,0}_{NS}^{\mathbb{Z}_4}+i\Bket{0,1;0,0}_{NS}^{\mathbb{Z}_4}\Big]\nonumber\\
&\Bket{[2,-2,0]\otimes [2,2,0]}^\sigma &&\leftrightarrow&&\frac{1}{2}\Big[+i\,\Bket{1,0;1,0}_{NS}^{\mathbb{Z}_4}-i\,\Bket{0,1;0,1}_{NS}^{\mathbb{Z}_4}+\Bket{1,0;0,1}_{NS}^{\mathbb{Z}_4}+\Bket{0,1;1,0}_{NS}^{\mathbb{Z}_4}\Big]\nonumber\\
&\Bket{[2,2,0]\otimes [2,-2,0]}^\sigma &&\leftrightarrow&&\frac{1}{2}\Big[ -i\,\Bket{1,0;1,0}_{NS}^{\mathbb{Z}_4}+i\,\Bket{0,1;0,1}_{NS}^{\mathbb{Z}_4}+\Bket{1,0;0,1}_{NS}^{\mathbb{Z}_4}+\Bket{0,1;1,0}_{NS}^{\mathbb{Z}_4}\Big]\nonumber\\
&\Bket{[2,0,0]\otimes [2,0,0]}^\sigma &&\leftrightarrow&&  \frac{1}{2}\Big[\Bket{2,0;0,0}_{NS}^{\mathbb{Z}_4}+\Bket{1,1;-1,1}_{NS}^{\mathbb{Z}_4}+\Bket{0,2;0,0}_{NS}^{\mathbb{Z}_4}+\Bket{1,1;1,-1}_{NS}^{\mathbb{Z}_4}\Big]\nonumber\,.
\end{align}
We use the notation $\Bket{m_1,w_1;m_2,w_2}$ for the Ishibashi state corresponding to the primary field $V_{m_1,w_1;m_2;w_2}$ (see definition in \eqref{Vpf} in the appendix).
The remaining NS Ishibashi states that appear in \eqref{pbs} correspond to $G^+_{1,2}$ descendants of these. (The $T^2/\mathbb{Z}_4$ Ishibashi states are defined so that they already include the sum over $s_1,s_2$). 

From experience with boundary states in (orbifolds of) flat space, we expect to see the boundary state (in the unstwisted sector, in the case of an orbifold) written as an infinite sum of coherent states labeled by winding and momentum quantum numbers, with a weight dependent on position and Wilson line. The boundary states that will result from the analysis of this section can be written in this form, but from the point of view of the extended chiral algebra that includes the $U(1)$  current $J$ it is more natural to consider a finite number of primary fields and hence Ishibashi states. The sum over winding and momentum arises from the sum over $J\bar{J}$-descendants in the  Ishibashi states.

In the R sector we see the couplings to the ground states obtained by spectral flow from the $(c,c)$ states ($(a,c)$ from the point of view of $T^2/\mathbb{Z}_4$):
\begin{align}\label{RgsIsh}
&\Bket{[0,1,1]\otimes [0,-1,-1]}^\sigma& &\leftrightarrow&& \frac{e^{i\pi/4}}{2}\big[-i\Bket{\sigma^{(\frac{2}{4})}_{00}}_{R}+i\Bket{\sigma^{(\frac{2}{4})}_{11}}_{R}+\sqrt{2}\Bket{\sigma^{(\frac{2}{4})}_{01}}_{R}\big]\nonumber\\
&\Bket{[1,2,1]\otimes [1,-2,-1]}^\sigma&&\leftrightarrow&& \frac{e^{i\pi/2}}{\sqrt{2}}\big[\Bket{\sigma^{(\frac{2}{4})}_{00}}_{R}+\Bket{\sigma^{(\frac{2}{4})}_{11}}_{R}\big]\\
&\Bket{[2,3,1]\otimes [2,-3,-1]}^\sigma& &\leftrightarrow &&\frac{e^{i3\pi/4}}{2}\big[+i\Bket{\sigma^{(\frac{2}{4})}_{00}}_{R}-i\Bket{\sigma^{(\frac{2}{4})}_{11}}_{R}+\sqrt{2}\Bket{\sigma^{(\frac{2}{4})}_{01}}_{R}\big]\nonumber\,.
\end{align}
In addition, there are some couplings to excited twisted sector fields (see \eqref{tau}  in the appendix):
\begin{align}\label{RIsh}
&\Bket{[0,-3,-1]\otimes [0,3,-1]}^\sigma& &\leftrightarrow&&\frac{e^{i3\pi/4}}{2}\big[+i\Bket{\tau^{(\frac{2}{4})}_{00}}_{R}-i\Bket{\tau^{(\frac{2}{4})}_{11}}_{R}+\sqrt{2}\Bket{\tau^{(\frac{2}{4})}_{01}}_{R}\big]\nonumber\\
&\Bket{[1,0,1]\otimes [1,0,1]}^\sigma &&\leftrightarrow&&\frac{e^{-i\pi/2}}{\sqrt{2}}\big[\Bket{\sigma^{-(\frac{2}{4})}_{00}}_{R}+\Bket{\sigma^{-(\frac{2}{4})}_{11}}_{R}\big]\\
&\Bket{[2,-1,1]\otimes [2,1,1]}^\sigma &&\leftrightarrow&&\frac{e^{i\pi/4}}{2}\big[-i\Bket{\tau^{(\frac{2}{4})}_{00}}_{R}+i\Bket{\tau^{(\frac{2}{4})}_{11}}_{R}+\sqrt{2}\Bket{\tau^{(\frac{2}{4})}_{01}}_{R}\big]\,.\nonumber
\end{align}

We can use this dictionary to translate the permutation boundary states \eqref{pbs} into $T^2/\mathbb{Z}_4$ boundary states. As an example, let us consider the cycle $(a)$ in Figure \ref{branes}. The relations \eqref{amap} and \eqref{mapkM} tell us that a brane wrapping this cycle is described by the permutation boundary state  $|[0,M,0,0]\rangle\,$, with $M=0,2$ for the two allowed values of the Wilson line. 
The relations in  eq. \eqref{NSIsh}-\eqref{RIsh} allow us to rewrite this boundary states \eqref{pbs} as
\begin{equation}\label{pbsT}
\begin{split}
|[0,&\{M=0,2\},0,0]\rangle= \\\frac{1}{2\sqrt{2}}\Big[&\Bket{0,0;0,0}_{NS}+\Bket{2,0;0,0}_{NS}^{\mathbb{Z}_4}+\Bket{1,1;-1,1}_{NS}^{\mathbb{Z}_4}+\Bket{0,2;0,0}_{NS}^{\mathbb{Z}_4}+\Bket{1,1;1,-1}_{NS}^{\mathbb{Z}_4}\\&+\Bket{1,0;0,0}_{NS}^{\mathbb{Z}_4}+(-1)^{\frac{M}{2}}\,\Bket{0,1;0,0}_{NS}^{\mathbb{Z}_4}+(-1)^{\frac{M}{2}}\,\Bket{1,0;0,1}_{NS}^{\mathbb{Z}_4}+(-1)^{\frac{M}{2}}\,\Bket{0,1;1,0}_{NS}^{\mathbb{Z}_4}\\
& +i\,\Bket{\tau_{01}^{(\frac{2}{4})}}_R-i\frac{(-1)^{\frac{M}{2}}}{\sqrt{2}}\,\Bket{\tau^{(\frac{2}{4})}_{0}}_{R}+i\frac{(-1)^{\frac{M}{2}}}{\sqrt{2}}\,\Bket{\tau^{(\frac{2}{4})}_{1}}_{R}-i\frac{(-1)^{\frac{M}{2}}}{\sqrt{2}}\,\Bket{\sigma^{-(\frac{2}{4})}_{0}}_{R}-i\frac{(-1)^{\frac{M}{2}}}{\sqrt{2}}\,\Bket{\sigma^{-(\frac{2}{4})}_{1}}_{R}\\&+i\,\Bket{\sigma_{01}^{(\frac{2}{4})}}_R+i(-1)^{\frac{M}{2}}\sqrt{2}\,\Bket{\sigma_{0}^{(\frac{2}{4})}}_R\Big]
\end{split}
\end{equation} 
where now on the right-hand sides we have linear combinations of $T^2/\mathbb{Z}_4$ Ishibashi states.   The topological one-point functions are encoded in the Ramond terms that appear in the last line and note that they match the ansatz \eqref{amap}, that was motivated by spacetime considerations in section \ref{Abranes}. In the NS sector there are only untwisted sector contributions. If we substitute $\theta\equiv\frac{M}{2}\pi\,$, the boundary state takes the right form for  a brane with a Wilson line $\theta\,$: in particular if $\theta=\pi$  we see that a negative sign appears in front of the terms that contain the sum over odd winding sectors. 

Similarly, we can write the boundary state for a brane wrapping the cycle $(b)$ in Figure \ref{branes}. In this case we need to take the permutation boundary states with $L=0$ and $M=-2,-4\,$, corresponding to the two choices of Wilson line. Repeating the steps that led to \eqref{pbsT}, we get
\begin{equation}
\begin{split}
|[0,&\{M=-2,-4\},0,0]\rangle= \\\frac{1}{2\sqrt{2}}\Big[&\Bket{0,0;0,0}_{NS}+\Bket{2,0;0,0}_{NS}^{\mathbb{Z}_4}+\Bket{1,1;-1,1}_{NS}^{\mathbb{Z}_4}+\Bket{0,2;0,0}_{NS}^{\mathbb{Z}_4}+\Bket{1,1;1,-1}_{NS}^{\mathbb{Z}_4}\\&-\Bket{1,0;0,0}_{NS}^{\mathbb{Z}_4}+(-1)^{\frac{M}{2}+1}\,\Bket{0,1;0,0}_{NS}^{\mathbb{Z}_4}+(-1)^{\frac{M}{2}+1}\,\Bket{1,0;0,1}_{NS}^{\mathbb{Z}_4}+(-1)^{\frac{M}{2}+1}\,\Bket{0,1;1,0}_{NS}^{\mathbb{Z}_4}\\
& -i\,\Bket{\tau_{01}^{(\frac{2}{4})}}_R+i\frac{(-1)^{\frac{M}{2}}}{\sqrt{2}}\,\Bket{\tau^{(\frac{2}{4})}_{0}}_{R}-i\frac{(-1)^{\frac{M}{2}}}{\sqrt{2}}\,\Bket{\tau^{(\frac{2}{4})}_{1}}_{R}+i\frac{(-1)^{\frac{M}{2}}}{\sqrt{2}}\,\Bket{\sigma^{-(\frac{2}{4})}_{0}}_{R}+i\frac{(-1)^{\frac{M}{2}}}{\sqrt{2}}\,\Bket{\sigma^{-(\frac{2}{4})}_{1}}_{R}\\&-i\,\Bket{\sigma_{01}^{(\frac{2}{4})}}_R+i(-1)^{\frac{M}{2}}\sqrt{2}\,\Bket{\sigma_{0}^{(\frac{2}{4})}}_R\Big]
\end{split}
\end{equation}
Again, note that this is consistent with the identification made in section \ref{Abranes}. The negative sign in front of terms that contain a sum over odd momenta arises because the cycle $(b)$ is displaced by $\pi R$ with respect to cycle $(a)\,$. 

\subsection{Tensor product boundary states}\label{MMTB}

Let us now come back to the tensor product branes that were discussed briefly in section \ref{LGbranes}. We know from the computation of one-point-functions in the topological theory that these branes are not charged under twisted sector fields. Moreover they have moduli, so they should be identified with ($\mathbb{Z}_4$-orbits of) D1-branes on $T^2/\mathbb{Z}_4$ that are not constrained to the orbifold fixed points. 

The boundary states that describe these branes in the minimal model framework are known \cite{Brunner, Ashok,BGpermutation}. They have the form
\begin{equation}\label{tbs}
|[L_1,S_1,L_2,S_2]\rangle = \sum_{l_1,s_1,l_2,s_2} C^{L_1,S_1,L_2,S_2}_{l_1,s_1,l_2,s_2}\,\Bket{[l_1,0,s_1]\otimes [l,0, s_2]}^{\sigma} \,,
\end{equation}
with 
\begin{multline}
C^{L_1,S_1,L_2,S_2}_{l_1,s_1,l_2,s_2}=\frac{1}{2\sqrt{2}}\left(e^{-i\pi \frac{S_1 s_1}{2}}\,\frac{\sin[\frac{\pi}{4} (L_1+1)(l_1+1)] }{\sqrt{\sin[\frac{\pi}{4}(l_1+1)]}} +e^{-i\pi \frac{(S_1+2) s_1}{2}}\,\frac{\sin[\frac{\pi}{4} (3-L_1)(l_1+1)] }{\sqrt{\sin[\frac{\pi}{4}(l_1+1)]}}  \right)\\ 
\times \left(e^{-i\pi \frac{S_2 s_2}{2}}\,\frac{\sin[\frac{\pi}{4} (L_2+1)(l_2+1)] }{\sqrt{\sin[\frac{\pi}{4}(l_2+1)]}} +e^{-i\pi \frac{(S_2+2) s_2}{2}}\,\frac{\sin[\frac{\pi}{4} (3-L_2)(l_2+1)] }{\sqrt{\sin[\frac{\pi}{4}(l_2+1)]}}  \right)\,.
\end{multline}
In \eqref{tbs} the sum runs over $l_i+s_i \in 2\mathbb{Z}\,$. The equivalence relation has been used to set $m_1=m_2=0\,$, so we sum over all values of $s_1,s_2\,$. As for the permutation boundary states, we set $S_1=S_2=0$ for consistency with the LG boundary conditions. The map between these boundary states and the tensor product factorizations is \cite{Ashok,BGpermutation}
\begin{equation}
|[L_1,0,L_2,0]\rangle \qquad \leftrightarrow \qquad (F_1=Y_1^{L_1+1})\otimes (F_2=Y_2^{L_2+1})\,,
\end{equation}
where on the right hand side the notation refers to the tensor product of LG factorizations, mentioned in section \ref{LGbranes}. 

The boundary state \eqref{tbs} satisfies the B-type boundary conditions
\begin{align}\label{tensbc}
&(T_1-\bar{T}_1)|B\rangle=0 &&(T_2-\bar{T}_2)|B\rangle=0\nonumber\\
&( G_1^\pm +i\eta\, \bar{G}_1^\pm)|B\rangle=0&&( G_2^\pm +i\eta\, \bar{G}_2^\pm)|B\rangle=0\\
& ( J_1+ \bar{J}_1)|B\rangle=0&&
 ( J_2+ \bar{J}_2)|B\rangle=0  \nonumber \ ,
\end{align}
which written in terms of $T^2/\mathbb{Z}_4$ currents become
\begin{align}
&(T-\bar{T})|B\rangle=0 &&
( G^\pm +i\eta\, \bar{G}^\pm)|B\rangle=0\\
& ( Q- \bar{Q})|B\rangle=0&&
 ( J- \bar{J})|B\rangle=0  \nonumber \ .
\end{align}
Comparing these relations with \eqref{permbcT} we see, as anticipated, that the only difference between the two is in the boundary condition along the circle associated with the $U(1)$ currents $J$ and $\bar{J}\,$.  

Let us look in more detail at the boundary state with $L_1=L_2=0\,$. This corresponds to the matrix factorization \eqref{tensfact} which, as was pointed out earlier, generates all other tensor product branes as bound states. As we did before, using the dictionary given in Table \ref{MMtoT} in the appendix,  we can translate the corresponding boundary state \eqref{tbs} into a boundary state on $T^2/\mathbb{Z}_4\,$  :
\begin{equation}\label{tbsT}
\begin{split}
|L_1&=0,0,L_2=0,0]\rangle = \Bket{0,0;0,0}_{NS}+\Bket{0,1;0,1}_{NS}^{\mathbb{Z}_4} +\Bket{1,0;1,0}_{NS}^{\mathbb{Z}_4} \\&+\frac{1}{2}\big[\Bket{2,0;0,0}_{NS}^{\mathbb{Z}_4}+\Bket{1,1;-1,1}_{NS}^{\mathbb{Z}_4}+\Bket{0,2;0,0}_{NS}^{\mathbb{Z}_4}+\Bket{1,1;1,-1}_{NS}^{\mathbb{Z}_4}\big]\\&+\sqrt{2}\,\big[\Bket{\sigma^{-(\frac{2}{4})}_{0}}_R+\Bket{\sigma^{-(\frac{2}{4})}_{1}}_R\big]  \,.
\end{split}
\end{equation}
Not all winding and momentum sectors appear in this case: for example, the sectors with one unit of momentum or one unit of winding are not included. The reason is that the corresponding Ishibashi states do not respect the boundary condition  on $J\,$. The boundary state \eqref{tbs} is thus only consistent at the self-dual radius. There is a twisted sector contribution in the Ramond sector, but since the corresponding primary field is not a ground state the brane does not carry twisted sector charge and it is therefore not constrained to the fixed points. Finally, note that the normalization is consistent with the conjecture of section \ref{LGbranes} that this boundary state should describe the superposition of two branes\footnote{ A factor of $\sqrt{2}$ relative to \eqref{pbs} comes from the the fact that this brane is extended along the diagonal of the square torus, as shown from the pattern of winding and momentum quantum numbers that appear in \eqref{tbsT}.}.

\begin{appendix}

\renewcommand{\thetable}{\thesection.\arabic{table}}
\setcounter{table}{0}

\section{Details of the map between $A_2\otimes A_2$ and $T^2/\mathbb{Z}_4$ primary fields}\label{map}

This appendix contains a summary of the map between the primary fields of $A_2\otimes A_2$ and $T^2/\mathbb{Z}_4\,$. Some facts about $\mathcal{N}=2$  minimal models are stated when needed; for a more complete review see \cite{Gepner, Qiu, MMS}.

The primary fields of a single $A_2$ minimal model, with their quantum numbers and representation in terms of free fields, are listed in Table \ref{oneMM}. They are labeled by $(l,m,s)\,$, with $l=0,1,2\,$, $m=-4,\ldots, +3$ (mod $8$), $s=0,\pm 1,2$ (mod  $4$). The labels are defined up to the equivalence relation $(l,m,s)\sim(2-l, m+4,s+2)\,$. In the NS sector $s\in 2\mathbb{Z}$ and in the R sector $s\in 2\mathbb{Z}+1\,$. Within each sector the quantum number $s$ serves the purpose of splitting the conformal family into two parts carrying opposite  $(-1)^F$ eigenvalues; states with different $s$ are reached by acting with $G^+$ (or $G^-$), with fermion number $(-1)^F=-1\,$.
A state labeled by $(l,m,s)$ has dimension and $U(1)$ charge given by
\begin{equation}
\begin{split}
h_{(l,m,s)}&=\frac{l(l+2)-m^2}{16}+\frac{s^2}{8}\phantom{+\frac{m-s-l}{2}}\qquad\qquad\text{if}\qquad -l\leq (m-s) \leq l\\
&=\frac{l(l+2)-m^2}{16}+\frac{s^2}{8} +\frac{m-s-l}{2} \qquad\qquad \text{if}\qquad l\leq(m-s)\leq 4-l\\
q_{(l,m,s)}&=\frac{m}{4}-\frac{s}{2}\qquad\qquad (\text{mod}\ 2)\ .
\end{split}
\end{equation}
The chiral primaries are those with quantum numbers $(l,l,0)\sim (2-l,l+4, 2)\,$. The spectral flow operator is $e^{-\frac{i}{2\sqrt{2}}\phi}\,$, with quantum numbers $(0,1,1)\,$, so  under spectral flow $(l,m,s)\,\rightarrow\, (l,m+1,s+1)\,$. This means that   the Ramond ground states   related to the chiral primaries by spectral flow have quantum numbers $(l,l+1,1)\sim (2-l,l-3,-1)\,$.

The primary fields of $A_2\otimes A_2 $ are obtained taking products of primary fields of the two minimal models. The NS primaries are listed in  Table \ref{MMxMM}.

The $A_2\otimes A_2$ primaries are in one-to-one correspondence with the primaries of the $T^2/\mathbb{Z}_4$ CFT, as a consequence of the map between the conformal algebra of the two models described in section \ref{CFT}. For the NS sector the correspondence was worked out in \cite{Chun}, but for the boundary state discussion of section \ref{MMBS} the map needs to be slightly refined. We adopt the notation
\begin{equation}\label{Vpf}
V_{m_1,w_1;m_2,w_2}=e^{\frac{i}{\sqrt{2}}(m_1 x^1+ m_2 x^2+ w_1 \tilde{x}^1+w_2 \tilde{x}^2)}
\end{equation}
for a primary field that creates a state with $m_i$ units of momentum ($i=1,2$ labels the directions  on $T^2$) and $w_i$ units of winding. Here $x^i=x^i_L+x^i_R$ and $\tilde{x}^i=x^i_L-x^i_R\,$. We denote by $e^{ ikB/4} \sigma^{(\frac{k}{4})}$ the primary field that creates the vacuum in the $k$-th twisted sector. See section \ref{CFT} and Figure \ref{twistfields} for a brief discussion of the twisted sectors. The twist fields associated with an inverse $\mathbb{Z}_4$ rotation are denoted by $e^{ -ikB/4} \sigma^{-(\frac{k}{4})}\,$. In addition to these, the list of primary fields includes some excited twist fields $e^{ \pm ikB/4}\tau^{\pm (\frac{k}{4})}\,$, defined by \cite{Dixon}
\begin{equation}\label{tau}
\partial x^+(z)\, \sigma^{\pm(\frac{k}{4})}(0) \sim z^{k/4-1}\tau^{\pm (\frac{k}{4})}(0)+\,\ldots\ .
\end{equation}

With these conventions, the primary fields of $T^2/\mathbb{Z}_4$ in the NS sector are those listed in Table \ref{pfT}. Note that the list contains only a finite subset of all the vertex operators that create winding and momentum sectors. All other $\mathbb{Z}_4$-invariant vertex operators are obtained from those in this list by acting with the $U(1)$ current \eqref{Uone} associated with the enhanced symmetry at the self-dual radius, so with respect to the enlarged chiral algebra that includes $J$ (and $\bar{J}$) they can be treated as descendants. 

As explained in section \ref{CFT}, the operators of the minimal model map to linear combination of $T^2/\mathbb{Z}_4$ operators with definite $U(1)$ charge, but there is a small technical point that should be mentioned for clarity. There is a phase ambiguity, since each eigenvector of the $U(1)$ charge matrix can be multiplied by an arbitrary phase. This phase becomes significant when one tries to compare the results for various boundary observables with the minimal models/LG side. The map between LG variables and $T^2/\mathbb{Z}_4$ chiral primaries is given in section \ref{CFT}. However, comparing the results for the disc one-point functions one finds that they match only upon introducing some nontrivial phases in the definition of the $A_2\otimes A_2$ Ishibashi states (see section \ref{MMBS}). Note that this does not affect any overlap computations. After fixing these phases for the chiral primaries to match the LG results, the ambiguity is fixed for all the other primary fields, simply requiring consistency of the OPE's. A different choice of phases would presumably give a basis of branes different from the one picked out by the Landau-Ginzburg factorizations.

Taking into account the few subtleties just mentioned, the map that we adopt to relate the $A_2\otimes A_2$ quantum numbers to the $T^2/\mathbb{Z}_4$ primary fields in the NS sector is given in Table \ref{MMtoT}. It is also important to know the map for the Ramond ground states, since these are related by spectral flow to the chiral primary fields. This map is given in Table \ref{RRgs}.

\begin{table}
\begin{center}
\begin{tabular}{|l|c|c|c|}\hline
\multirow{2}{*}{$(l,m,s)$} & \multirow{2}{*}{$h$}&\multirow{2}{*}{$ q$}&Free Field \\ &&&Representation\\\hline
$(0,0,0)\sim(2,-4,2)\phantom{\big(}$& $0$ & $0$&$\II$\\
$(0,-4,0)\sim (2,0,2) \phantom{\big(}$& $1$& $1$&$e^{i\sqrt{2}\phi}$\\
$(0,-2,0)\sim (2,2,2)\phantom{\big(}$& $\frac{3}{4}$& $-\frac{1}{2} $&$\varepsilon\, e^{-\frac{i}{\sqrt{2}}\phi}$\\
$(0,2,0)\sim(2,-2,2) \phantom{\big(}$&$ \frac{3}{4}$&$ \frac{1}{2}$&$\varepsilon\, e^{\frac{i}{\sqrt{2}}\phi}$\\
$(0,0,2)\sim(2,-4,0)\phantom{\big(}$&$\frac{3}{2}$&$1$&$\varepsilon\,e^{i\sqrt{2}\phi}$\\
$(0,-4,2)\sim(2,0,0)\phantom{\big(}$&$\frac{1}{2}$&$0$&$\varepsilon$\\
$(0,-2,2)\sim(2,2,0)\phantom{\big(}$&$\frac{1}{4}$&$\frac{1}{2}$&$e^{\frac{i}{\sqrt{2}}\phi}$\\
$(0,2,2)\sim(2,-2,0)\phantom{\big(}$&$\frac{1}{4}$&$-\frac{1}{2}$&$e^{-\frac{i}{\sqrt{2}}\phi}$\\
$(0,-3,-1)\sim(2,1,1)\phantom{\big(}$&$\frac{9}{16}$&$-\frac{1}{4}$&$\varepsilon\,e^{-\frac{i}{2\sqrt{2}}\phi}$\\
$(0,-1,-1)\sim(2,3,1)\phantom{\big(}$&$\frac{1}{16}$&$\frac{1}{4}$&$e^{\frac{i}{2\sqrt{2}}\phi}$\\
$(0,1,-1)\sim(2,-3,1)\phantom{\big(}$&$1+\frac{1}{16}$&$\frac{3}{4}$&$\varepsilon\,e^{\frac{i3}{2\sqrt{2}}\phi}$\\
$(0,3,-1)\sim(2,-1,1)\phantom{\big(}$&$\frac{9}{16}$&$-\frac{3}{4}$&$e^{-\frac{i3}{2\sqrt{2}}\phi}$\\
$(0,-3,1)\sim(2,1,-1)\phantom{\big(}$&$\frac{9}{16}$&$\frac{3}{4}$&$e^{\frac{i3}{2\sqrt{2}}\phi}$\\
$(0,-1,1)\sim(2,3,-1)\phantom{\big(}$&$1+\frac{1}{16}$&$-\frac{3}{4}$&$\varepsilon\,e^{-\frac{i3}{2\sqrt{2}}\phi}$\\
$(0,1,1)\sim(2,-3,-1)\phantom{\big(}$&$\frac{1}{16}$&$-\frac{1}{4}$&$e^{-\frac{i}{2\sqrt{2}}\phi}$\\
$(0,3,1)\sim(2,-1,-1)\phantom{\big(}$&$\frac{9}{16}$&$\frac{1}{4}$&$\varepsilon\,e^{\frac{i}{2\sqrt{2}}\phi}$\\
$(1,-3,0)\sim(1,1,2)\phantom{\big(}$&$\frac{5}{8}$&$-\frac{3}{4}$&$\sigma\, e^{-\frac{i3}{2\sqrt{2}}\phi}$\\
$(1,-1,0)\sim(1,3,2)\phantom{\big(}$&$\frac{1}{8}$&$-\frac{1}{4}$&$\sigma\, e^{-\frac{i}{2\sqrt{2}}\phi}$\\
$(1,1,0)\sim(1,-3,2)\phantom{\big(}$&$\frac{1}{8}$&$\frac{1}{4}$&$\sigma\, e^{\frac{i}{2\sqrt{2}}\phi}$\\
$(1,3,0)\sim(1,-1,2)\phantom{\big(}$&$\frac{5}{8}$&$\frac{3}{4}$&$\sigma\, e^{\frac{i3}{2\sqrt{2}}\phi}$\\
$(1,0,-1)\sim(1,-4,1)\phantom{\big(}$&$\frac{5}{16}$&$\frac{1}{2}$&$\sigma\, e^{\frac{i}{\sqrt{2}}\phi}$\\
$(1,-2,-1)\sim(1,2,1)\phantom{\big(}$&$\frac{1}{16}$&$0$&$\sigma$\\
$(1,0,1)\sim(1,-4,-1)\phantom{\big(}$&$\frac{5}{16}$&$-\frac{1}{2}$&$\sigma\, e^{-\frac{i}{\sqrt{2}}\phi}$\\
$(1,-2,1)\sim(1,2,-1)\phantom{\big(}$&$1+\frac{1}{16}$&$1$&$\sigma\,e^{i\sqrt{2}\phi}$\\\hline
\end{tabular}\caption{Primary fields of a single $A_2$ minimal model. For every allowed set of quantum numbers $(l,m,s)\sim(2-l, m+4,s+2)$ we have listed the corresponding conformal dimension and $U(1)$ charge (it is understood that $\bar{h}=h$ and $\bar{q}=q\,$). The free field representation is written in terms of a free boson $\phi$ and, for the fermionic component, the Ising variables $\sigma$ and $\varepsilon$.  Note that the labels $(0,0,2) $ and $(0,1,1)$ correspond, respectively, to the supercurrent $G^+$ (since $\varepsilon=\psi\bar{\psi}$) and the spectral flow operator. }\label{oneMM}
\end{center}
\end{table}

\begin{table}
\begin{center}
\begin{tabular}{|c|l|l|c|}\hline
\multirow{2}{*}{$(h,q)$}&\multirow{2}{*}{$(l_1,m_1,s_1)$ }&\multirow{2}{*}{$(l_2,m_2,s_2)$}&Free Field \\ &&&Representation\\\hline
$(0,0)$&$(0,0,0)$&$(0,0,0)$&$\II$\\
$(\frac{1}{4},0)$&$(1,1,0)$&$(1,-1,0)$&$\sigma_1 e^{+\frac{i}{2\sqrt{2}}\phi_1}\,\sigma_2 e^{-\frac{i}{2\sqrt{2}}\phi_2}$\\
&$(1,-1,0)$&$(1,1,0)$&$\sigma_1 e^{-\frac{i}{2\sqrt{2}}\phi_1}\,\sigma_2 e^{+\frac{i}{2\sqrt{2}}\phi_2}$\\
$(\frac{1}{2},0)$&$(2,0,0)$&$(0,0,0)$&$\varepsilon_1$\\
&$(0,0,0)$&$(2,0,0)$&$\varepsilon_2$\\
&$(2,2,0)$&$(2,-2,0)$&$e^{+\frac{i}{\sqrt{2}}\phi_1}\,e^{-\frac{i}{\sqrt{2}}\phi_2}$\\
&$(2,-2,0)$&$(2,2,0)$&$e^{-\frac{i}{\sqrt{2}}\phi_1}\,e^{+\frac{i}{\sqrt{2}}\phi_2}$\\
$(1,0)$&$(2,0,0)$&$(2,0,0)$&$\varepsilon_1\,\varepsilon_2$\\\hline
$(\frac{1}{8},\pm\frac{1}{4})$&$(1,\pm1,0)$&$(0,0,0)$&$\sigma_1 e^{\pm\frac{i}{2\sqrt{2}}\phi_1}$\\
&$(0,0,0)$&$(1,\pm1,0)$&$\sigma_2 e^{\pm\frac{i}{2\sqrt{2}}\phi_2}$\\
$(\frac{1}{4},\pm\frac{1}{2})$&$(2,\pm2,0)$&$(0,0,0)$&$e^{\pm\frac{i}{\sqrt{2}}\phi_1}$\\
&$(0,0,0)$&$(2,\pm2,0)$&$e^{\pm\frac{i}{\sqrt{2}}\phi_2}$\\
&$(1,\pm1,0)$&$(1,\pm1,0)$&$\sigma_1 e^{\pm\frac{i}{2\sqrt{2}}\phi_1}\,\sigma_2 e^{\pm\frac{i}{2\sqrt{2}}\phi_2}$\\
$(\frac{3}{8},\pm\frac{3}{4})$&$(1,\pm1,0)$&$(2,\pm2,0)$&$\sigma_1 e^{\pm\frac{i}{2\sqrt{2}}\phi_1}\,e^{\pm\frac{i}{\sqrt{2}}\phi_2}$\\
&$(2,\pm2,0)$&$(1,\pm1,0)$&$e^{\pm\frac{i}{\sqrt{2}}\phi_1}\,\sigma_2 e^{+\pm\frac{i}{2\sqrt{2}}\phi_2}$\\
$(\frac{1}{2},\pm1)$&$(2,\pm2,0)$&$(2,\pm2,0)$&$e^{\pm\frac{i}{\sqrt{2}}\phi_1}\,e^{\pm\frac{i}{\sqrt{2}}\phi_2}$\\\hline
$(\frac{3}{8},\pm\frac{1}{4})$&$(1,\mp1,0)$&$(2,\pm2,0)$&$\sigma_1 e^{\mp\frac{i}{2\sqrt{2}}\phi_1}\,e^{\pm\frac{i}{\sqrt{2}}\phi_2}$\\
&$(2,\pm2,0)$&$(1,\mp1,0)$&$e^{\pm\frac{i}{\sqrt{2}}\phi_1}\,\sigma_2 e^{\mp\frac{i}{2\sqrt{2}}\phi_2}$\\
$(\frac{5}{8},\pm\frac{1}{4})$&$(1,\pm1,0)$&$(2,0,0)$&$\sigma_1 e^{\pm\frac{i}{2\sqrt{2}}\phi_1}\,\varepsilon_2$\\
&$(2,0,0)$&$(1,\pm1,0)$&$\varepsilon_1\,\sigma_2 e^{\pm\frac{i}{2\sqrt{2}}\phi_2}$\\
$(\frac{3}{4},\pm\frac{1}{2})$&$(2,\pm2,0)$&$(2,0,0)$&$e^{\pm\frac{i}{\sqrt{2}}\phi_1}\,\varepsilon_2$\\
&$(2,0,0)$&$(2,\pm2,0)$&$\varepsilon_1\,e^{\pm\frac{i}{\sqrt{2}}\phi_2}$\\\hline
\end{tabular}\caption{Primary fields of $A_2\otimes A_2$ in the NS sector. Only the fields with $s_1=s_2=0$ have been included: the remaining fields are obtained from these acting with the supercurrents $G^+_{1,2}\,$. Those listed between the two horizontal lines are the chiral/anti-chiral primaries, depending on the sign of $q\,$.}\label{MMxMM}
\end{center}
\end{table}

\begin{table}
\begin{center}
\begin{tabular}{|l|c|l|}\hline
$\phantom{\Big(}$&$(h,q)$&$T^2/\mathbb{Z}_4$ Vertex Operator\\\hline
untwisted sector:$\phantom{\Big(}$ &$(\frac{1}{4},0)$&$V^{\mathbb{Z}_4}_{1,0;0,0}\equiv\frac{1}{2}(V_{1,0;0,0}+V_{0,0;1,0}+V_{-1,0;0,0}+V_{0,0;-1,0})$\\
$\phantom{\Big(}$&&$V^{\mathbb{Z}_4}_{0,1;0,0}\equiv\frac{1}{2}(V_{0,1;0,0}+V_{0,0;0,1}+V_{0,-1;0,0}+V_{0,0;0,-1})$\\
$\phantom{\Big(}$&$(\frac{1}{2},0)$&$V^{\mathbb{Z}_4}_{1,0;1,0}\equiv\frac{1}{2}(V_{1,0;1,0}+V_{-1,0;1,0}+V_{-1,0;-1,0}+V_{1,0;-1,0})$\\
$\phantom{\Big(}$&&$V^{\mathbb{Z}_4}_{1,0;0,1}\equiv\frac{1}{2}(V_{1,0;0,1}+V_{0,-1;1,0}+V_{-1,0;0,-1}+V_{0,1;-1,0})$\\
$\phantom{\Big(}$&&$V^{\mathbb{Z}_4}_{0,1;0,1}\equiv\frac{1}{2}(V_{0,1;0,1}+V_{0,-1;0,1}+V_{0,-1;0,-1}+V_{0,1;0,-1})$\\
$\phantom{\Big(}$&&$V^{\mathbb{Z}_4}_{0,1;1,0}\equiv\frac{1}{2}(V_{0,1;1,0}+V_{-1,0;0,1}+V_{0,-1;-1,0}+V_{1,0;0,-1})$\\
$\phantom{\Big(}$&$(1,0)$&$\frac{1}{2}(V^{\mathbb{Z}_4}_{2,0;0,0}+V^{\mathbb{Z}_4}_{1,1;-1,1}+V^{\mathbb{Z}_4}_{0,2;0,0}+V^{\mathbb{Z}_4}_{1,1;1,-1})$\\
$\phantom{\Big(}$&&$\equiv\frac{1}{4}\big[(V_{2,0;0,0}+V_{0,0;2,0}+V_{-2,0;0,0}+V_{0,0;-2,0})$\\
$\phantom{\Big(}$&&$+\,(V_{1,1;-1,1}+V_{1,-1;1,1}+V_{-1,-1;1,-1}+V_{-1,1;-1,-1})$\\
$\phantom{\Big(}$&&$+\,(V_{0,2;0,0}+V_{0,0;0,2}+V_{0,-2;0,0}+V_{0,0;0,-2})$\\
$\phantom{\Big(}$&&$+\,(V_{1,1;1,-1}+V_{-1,1;1,1}+V_{-1,-1;-1,1}+V_{1,-1;-1,-1})\big]$\\
$\phantom{\Big(}$&$(\frac{1}{2},\pm1)$&$\psi^\pm\bar{\psi}^\mp$\\\hline
twisted sectors: $\phantom{\Big(}$&$(\frac{1}{8},\pm\frac{1}{4})$&$e^{\pm iB/4}\sigma^{\pm(\frac{1}{4})}_0\,,\ e^{iB/4}\sigma^{\pm(\frac{1}{4})}_1$\\
$\phantom{\Big(}$&$(\frac{1}{4},\pm\frac{1}{2})$&$e^{\pm iB/2}\sigma^{\pm(\frac{2}{4})}_0\,,\ e^{\pm iB/2}\sigma^{\pm(\frac{2}{4})}_1\,,\ e^{\pm iB/2}\sigma^{\pm(\frac{2}{4})}_{01}$\\
$\phantom{\Big(}$&$(\frac{3}{8},\pm\frac{3}{4})$&$e^{\pm i3B/4}\sigma^{\pm (\frac{3}{4})}_0\,,\ e^{\pm i3B/4}\sigma^{\pm(\frac{3}{4})}_1$\\
\phantom{\Big(}&$(\frac{3}{8},\pm\frac{1}{4})$&$e^{iB/4}\tau^{\pm(\frac{1}{4})}_0\,,\ e^{iB/4}\tau^{\pm(\frac{1}{4})}_1$\\
\phantom{\Big(}&$(\frac{3}{4},\pm\frac{1}{2})$&$e^{iB/2}\tau^{\pm(\frac{2}{4})}_0\,,\ e^{iB/2}\tau^{\pm(\frac{2}{4})}_1\,,\ e^{iB/2}\tau^{\pm(\frac{2}{4})}_{01}$\\
\hline
\end{tabular}\caption{Primary fields of $T^2/\mathbb{Z}_4$ in the NS sector. Here the convention is $\bar{h}=h$ and $\bar{q}=-q\,$. We distinguish the $\mathbb{Z}_4$ orbit of an operator with the specified winding and momentum with a superscript $\mathbb{Z}_4\,$. 
The notation $\sigma^{\pm(\frac{k}{4})}$ denotes twist fields associated with rotations by $\omega^{\pm k}\,$, while $\tau^{\pm(\frac{k}{4})}$ are the excited twist  fields defined as in \eqref{tau}. Note that in fact the twist fields $e^{iB/2}\tau^{(\frac{2}{4})}$ are not independent primaries, since one can check that  $(\tau^{(\frac{2}{4})}_0+\tau^{(\frac{2}{4})}_1)\sim (G_1^++G^+_2) \cdot (\sigma^{-(\frac{2}{4})}_0+\sigma^{-(\frac{2}{4})}_1)\,$. }\label{pfT}
\end{center}
\end{table}

\begin{table}
\begin{center}
\begin{tabular}{|l|l|}\hline
$A_2\otimes A_2$ labels $\phantom{\Big(}\qquad$& $T^2/\mathbb{Z}_4$ primary fields \\\hline
$(1,1,0)\otimes(1,-1,0)\phantom{\Big(}$& $\frac{1}{\sqrt{2}}e^{-i\pi/4}(V^{\mathbb{Z}_4}_{1,0;0,0}+ i V^{\mathbb{Z}_4}_{0,1;0,0})$\\
$(1,-1,0)\otimes(1,1,0)\phantom{\Big(}$&$\frac{1}{\sqrt{2}}e^{i\pi/4}(V^{\mathbb{Z}_4}_{1,0;0,0}- i V^{\mathbb{Z}_4}_{0,1;0,0})$\\
$(2,0,0)\otimes(0,0,0)\phantom{\Big(}$&$\frac{1}{2}(V^{\mathbb{Z}_4}_{1,0;1,0}+V^{\mathbb{Z}_4}_{0,1;0,1}+V^{\mathbb{Z}_4}_{1,0;0,1}- V^{\mathbb{Z}_4}_{0,1;1,0})$\\
$(0,0,0)\otimes(2,0,0)\phantom{\Big(}$&$\frac{1}{2}(V^{\mathbb{Z}_4}_{1,0;1,0}+V^{\mathbb{Z}_4}_{0,1;0,1}-V^{\mathbb{Z}_4}_{1,0;0,1}+V^{\mathbb{Z}_4}_{0,1;1,0})$\\
$(2,2,0)\otimes(2,-2,0)\phantom{\Big(}$&$\frac{1}{2}(+iV^{\mathbb{Z}_4}_{1,0;1,0}-iV^{\mathbb{Z}_4}_{0,1;0,1}+V^{\mathbb{Z}_4}_{1,0;0,1}+ V^{\mathbb{Z}_4}_{0,1;1,0})$\\
$(2,-2,0)\otimes(2,2,0)\phantom{\Big(}$&$\frac{1}{2}(-iV^{\mathbb{Z}_4}_{1,0;1,0}+iV^{\mathbb{Z}_4}_{0,1;0,1}+V^{\mathbb{Z}_4}_{1,0;0,1}+ V^{\mathbb{Z}_4}_{0,1;1,0})$\\
$(2,0,0)\otimes(2,0,0)\phantom{\Big(}$&$\frac{1}{2}(V^{\mathbb{Z}_4}_{2,0;0,0}+V^{\mathbb{Z}_4}_{1,1;-1,1}+V^{\mathbb{Z}_4}_{0,2;0,0}+V^{\mathbb{Z}_4}_{1,1;1,-1})$\\
$(1,\pm1,0)\otimes(0,0,0)\phantom{\Big(}$&$e^{\pm i3\pi/8}\,\frac{e^{\pm iB/4}}{\sqrt{2}}(\sigma^{\pm(\frac{1}{4})}_0-i \sigma^{\pm(\frac{1}{4})}_1)$\\
$(0,0,0)\otimes(1,\pm1,0)\phantom{\Big(}$&$e^{\pm i\pi/8}\,\frac{e^{\pm iB/4}}{\sqrt{2}}(-i\sigma^{\pm(\frac{1}{4})}_0+ \sigma^{\pm(\frac{1}{4})}_1)$\\
$(2,\pm2,0)\otimes(0,0,0)\phantom{\Big(}$&$e^{\pm i3\pi/4}\,\frac{e^{\pm iB/2}}{2}(+i\sigma^{\pm(\frac{2}{4})}_0-i\sigma^{\pm(\frac{2}{4})}_1+\sqrt{2}\sigma^{\pm(\frac{2}{4})}_{01})$\\
$(0,0,0)\otimes(2,\pm2,0)\phantom{\Big(}$&$e^{\pm i\pi/4}\,\frac{e^{\pm i B/2}}{2}(-i\sigma^{\pm(\frac{2}{4})}_0+i\sigma^{\pm(\frac{2}{4})}_1+\sqrt{2}\sigma^{\pm(\frac{2}{4})}_{01})$\\
$(1,\pm1,0)\otimes(1,\pm1,0)\phantom{\Big(}$&$e^{\pm i\pi/2}\,\frac{e^{\pm iB/2}}{\sqrt{2}}(\sigma^{\pm(\frac{2}{4})}_0+\sigma^{\pm(\frac{2}{4})}_1)$\\
$(1,\pm1,0)\otimes(2,\pm2,0)\phantom{\Big(}$&$e^{\mp i\pi/8}\,\frac{e^{\pm i3B/4}}{\sqrt{2}}(\sigma^{\pm(\frac{3}{4})}_0-i \sigma^{\pm(\frac{3}{4})}_1)$\\
$(2,\pm2,0)\otimes(1,\pm1,0)\phantom{\Big(}$&$e^{\mp i\pi/8}\,\frac{e^{\pm i3B/4}}{\sqrt{2}}(-i\sigma^{\pm(\frac{3}{4})}_0+ \sigma^{\pm(\frac{3}{4})}_1)$\\
$(2,\pm2,0)\otimes(2,\pm2,0)\phantom{\Big(}$&$\psi^\pm\bar{\psi}^\mp$\\
$(1,\mp1,0)\otimes(2,\pm2,0)\phantom{\Big(}$&$e^{\pm i3\pi/8}\,\frac{e^{\pm iB/4}}{\sqrt{2}}(\tau^{\pm(\frac{1}{4})}_0-i \tau^{\pm(\frac{1}{4})}_1)$\\
$(2,\pm2,0)\otimes(1,\mp1,0)\phantom{\Big(}$&$e^{\pm i\pi/8}\,\frac{e^{\pm iB/4}}{\sqrt{2}}(-i\tau^{\pm(\frac{1}{4})}_0+ \tau^{\pm(\frac{1}{4})}_1)$\\
$(2,\pm2,0)\otimes(2,0,0)\phantom{\Big(}$&$e^{\pm i3\pi/4}\,\frac{e^{\pm iB/2}}{2}(+i\tau^{\pm(\frac{2}{4})}_0-i\tau^{\pm(\frac{2}{4})}_1+\sqrt{2}\tau^{\pm(\frac{2}{4})}_{01})$\\
$(2,0,0)\otimes(2,\pm2,0)\phantom{\Big(}$&$e^{\pm i\pi/4}\,\frac{e^{\pm i B/2}}{2}(-i\tau^{\pm(\frac{2}{4})}_0+i\tau^{\pm(\frac{2}{4})}_1+\sqrt{2}\tau^{\pm(\frac{2}{4})}_{01})$\\
\hline
\end{tabular}\caption{Relation between $A_2\otimes A_2$ labels and $T^2/\mathbb{Z}_4$ primary fields in the NS sector.}\label{MMtoT}
\end{center}
\end{table}

\begin{table}
\begin{center}
\begin{tabular}{|l|r|c|l|}\hline
$(l_1,m_1,s_1)\otimes (l_2,m_2,s_2)\phantom{\Big(}$&$q$&Free Field Rep.&$T^2/\mathbb{Z}_4$ RR  Ground  States\\\hline
$(0,1,1)\otimes (0,1,1)\phantom{\Big(}$&$-\frac{1}{2}$&$e^{\frac{i}{2\sqrt{2}}(\phi_1+\phi_2)}$&$|0\rangle^{(0)}_{R,-}$\\
$(0,-1,-1)\otimes (0,-1,-1)\phantom{\Big(}$&$\frac{1}{2}$&$e^{-\frac{i}{2\sqrt{2}}(\phi_1+\phi_2)}$&$|0\rangle^{(0)}_{R,+}\equiv\psi_0^+\bar{\psi}_0^-|0\rangle_{R,-}$\\
$(1,2,1)\otimes (0,-1,-1)\phantom{\Big(}$&$\frac{1}{4}$&$\sigma_1\,e^{\frac{i}{2\sqrt{2}}\phi_2}$&$\frac{e^{i3\pi/8}}{\sqrt{2}}(\sigma^{(\frac{1}{4})}_0-i \sigma^{(\frac{1}{4})}_1)|0\rangle_{R}^{(\frac{1}{4})}$\\
$(0,-1,-1)\otimes (1,2,1)\phantom{\Big(}$&$\frac{1}{4}$&$\sigma_2\,e^{\frac{i}{2\sqrt{2}}\phi_1}$&$\frac{e^{i\pi/8}}{\sqrt{2}}( \sigma^{(\frac{1}{4})}_1-i\sigma^{(\frac{1}{4})}_0)|0\rangle_{R}^{(\frac{1}{4})}$\\
$(0,1,1)\otimes (0,-1,-1)\phantom{\Big(}$&$0$&$e^{\frac{i}{2\sqrt{2}}(\phi_1-\phi_2)}$&$\frac{e^{i\pi/4}}{\sqrt{2}}(-i\sigma^{(\frac{2}{4})}_0+i\sigma^{(\frac{2}{4})}_1+\sqrt{2}\sigma^{(\frac{2}{4})}_{01})|0\rangle_{R}^{(\frac{2}{4})}$\\
$(0,-1,-1)\otimes (0,1,1)\phantom{\Big(}$&$0$&$e^{\frac{i}{2\sqrt{2}}(\phi_2-\phi_1)}$&$\frac{e^{i3\pi/4}}{\sqrt{2}}(+i\sigma^{(\frac{2}{4})}_0-i\sigma^{(\frac{2}{4})}_1+\sqrt{2}\sigma^{(\frac{2}{4})}_{01})|0\rangle_{R}^{(\frac{2}{4})}$\\
$(1,2,-1)\otimes (1,2,1)\phantom{\Big(}$&$0$&$\sigma_1\sigma_2$&$\frac{i}{\sqrt{2}}(\sigma^{(\frac{2}{4})}_0+\sigma^{(\frac{2}{4})}_1)|0\rangle_{R}^{(\frac{2}{4})}$\\
$(0,1,1)\otimes (1,2,1)\phantom{\Big(}$&$-\frac{1}{4}$&$\sigma_2\,e^{-\frac{i}{2\sqrt{2}}\phi_1}$&$\frac{e^{-i\pi/8}}{\sqrt{2}}( \sigma^{(\frac{3}{4})}_1-i\sigma^{(\frac{3}{4})}_0)|0\rangle_{R}^{(\frac{3}{4})}$\\
$(1,2,1)\otimes (0,1,1)\phantom{\Big(}$&$-\frac{1}{4}$&$\sigma_1\,e^{-\frac{i}{2\sqrt{2}}\phi_2}$&$\frac{e^{-i\pi/8}}{\sqrt{2}}(\sigma^{(\frac{3}{4})}_0-i \sigma^{(\frac{3}{4})}_1)|0\rangle_{R}^{(\frac{3}{4})}$\\\hline
\end{tabular}\caption{Ramond ground states, related by spectral flow to the chiral primary fields in the NS sector.  The first column contains the minimal model labels and the free field representation is given in the third column. In the last column we have listed the corresponding ground states of the $T^2/\mathbb{Z}_4$ model. Note that they belong to the $(2-4q)$-th twisted sector, where $q$ is the $U(1)$ charge (second column). The superscript in $|0\rangle_{R}^{(\frac{k}{4})}$ signals that the state is created from the NS vacuum by the modified spin field $e^{i(\frac{k}{4}-\frac{1}{2})B}$ appropriate for the $k$-th twisted sector \cite{Dixon}, so that all the states listed above have the same conformal dimension. }\label{RRgs}
\end{center}
\end{table}

\end{appendix}

\end{document}